\documentclass[twocolumn]{aastex62}
\usepackage{amsmath}

\newcommand{\package}[1]{\textsl{#1}}

\newcommand{\msun}{\textrm{M}_\odot}

\newcommand{\kms}{\ensuremath{\textrm{km}~\textrm{s}^{-1}}}

\renewcommand{\tablename}{Table}

\shorttitle{Detecting cold streams}
\shortauthors{Pearson et al. }

\begin{document}\sloppy\sloppypar\raggedbottom\frenchspacing 

\title{Detecting Thin Stellar Streams in External Galaxies:\\ Resolved Stars \& Integrated Light}

 \author[0000-0003-0256-5446]{Sarah Pearson}
 \affiliation{Center for Computational Astrophysics, Flatiron Institute, 162 5th Av., New York City, NY 10010, USA}
 \email{spearson@flatironinstitute.org}
 \correspondingauthor{Sarah Pearson}

\author{Tjitske K. Starkenburg}
\affiliation{Center for Computational Astrophysics, Flatiron Institute, 162 5th Av., New York City, NY 10010, USA}

\author{Kathryn V. Johnston}
\affiliation{Department of Astronomy, Columbia University, Mail Code 5246, 550 West 120th Street, New York, New York 10027, USA}

\author{Benjamin F. Williams}
\affiliation{Department of Astronomy, Box 351580, University of Washington, Seattle, WA 98195, USA}

\author{Rodrigo A. Ibata}
\affiliation{Observatoire astronomique de Strasbourg, Université de Strasbourg, CNRS, UMR 7550, 11 rue de l’Université, F-67000 Strasbourg, France}
%
%

\begin{abstract}\noindent 
The morphology of thin stellar streams can be used to test the nature of dark matter. It is therefore crucial to extend searches for globular cluster streams to other galaxies than the Milky Way. In this paper, we investigate the current and future prospects of detecting globular cluster streams in external galaxies in resolved stars (e.g. with WFIRST) and using integrated light (e.g. with HSC, LSST and Euclid). In particular, we inject mock-streams to data from the PAndAS M31 survey, and produce simulated M31 backgrounds mimicking what WFIRST will observe in M31. Additionally, we estimate the distance limit to which globular cluster streams will be observable. Our results demonstrate that for a 1 hour (1000 sec.) exposure, using conservative estimates, WFIRST should detect globular cluster streams in resolved stars in galaxies out to distances of ${\sim}$3.5 Mpc (${\sim}$2 Mpc). This volume contains 199 (122) galaxies of which $>$90\% are dwarfs. With integrated light, thin streams can be resolved out to ${\sim}$100 Mpc with HSC and LSST and to ${\sim}$600 Mpc with WFIRST and Euclid. The low surface brightness of the streams (typically $>$30 mag/arcsec$^{2}$), however, will make them difficult to detect, unless the streams originate from very young clusters. We emphasize that if the external galaxies do not host spiral arms or galactic bars, gaps in their stellar streams provide an ideal test case for evidence of interactions with dark matter subhalos. Furthermore, obtaining a large samples of thin stellar streams can help constrain the orbital structure and hence the potentials of external halos. \end{abstract}

\section{Introduction} \label{sec:intro}
Stellar streams form when a gravitationally bound ensemble of stars tidally tears apart, due to an underlying galactic potential. To date, we have observed a multitude of stellar streams in our own Galaxy, emerging as leading and trailing arms from both open clusters (e.g. \citealt{roser19}) and globular clusters (e.g. GD1: \citealt{grillmair06}, Palomar 5: \citealt{oden01}), as well as dwarf galaxies (e.g. Sagittarius: \citealt{ibata01}, Orphan: \citealt{belokurov06}). Following the release of Gaia DR2 (\citealt{gaiadr2}), more than 60 stellar stream candidates have been suggested in the Milky Way alone (e.g. \citealt{ibata19}). 
Several stellar streams have additionally been discovered in external galaxies (e.g. \citealt{ibata00}, \citealt{delgado10}). Based on the widths and surface brightnesses of the stellar streams in external galaxies, these streams are likely relics from tidally disrupted dwarf galaxies (e.g. \citealt{delgado12}, \citealt{rich12}, \citealt{annibali12}).

Since the discovery of stellar streams, several studies have proposed to use the observed properties of streams to measure the mass distribution in our Galaxy, including its dark matter (e.g. \citealt{johnston99}, \citealt{koposov10}, \citealt{law10}, \citealt{sanderson15}, \citealt{bovy16}). Stellar streams from globular clusters (GCs) are of particular interest, as they are dynamically cold (i.e. the internal kinematics of globular clusters are much smaller than the clusters' orbital velocities around their host galaxies). As a consequence, the streams from GCs phase-mix slowly and leave behind stars moving coherently in phase-space along thin leading and trailing arm for several gigayears (Gyr) that can be dense enough to be detectable with today's surveys. 
Because the streams are so cold, they are particularly sensitive to any deviations from smooth, symmetric potentials - and hence particularly useful for probing dark matter distributions.

While studies to measure the potential of the Galaxy have typically relied on multiple dimensions of data, including kinematics, there are some specific examples where the morphology of thin streams are alone informative. For example, the $\Lambda$-cold dark matter ($\Lambda$CDM) paradigm predicts a specific distribution and mass range of dark matter subhalos in our Galaxy (see e.g. \citealt{diemand08}, \citealt{bovy17}, \citealt{bonaca19}). \citet{ibata02} and \citet{johnston02} showed that the interaction between dark matter subhalos can leave behind signatures in the structure of stellar streams. Density fluctuations in GC stellar streams can therefore, in principle, provide indirect evidence of interactions with dark matter substructure, and serve as a test of $\Lambda$CDM (e.g. \citealt{yoon11}, \citealt{erkal16}, \citealt{bovy17}, \citealt{bonaca19}).

In addition, the morphology of GC stellar streams provide broad constraints on  dark matter halo shapes, as only certain orbits in triaxial matter distributions allow thin, long streams to exist (\citealt{pearson15}). In particular, thin, long streams should only be detectable on regular or resonantly trapped orbits. Thus, their presence and location can provide a map of these regions in the orbit structure of a potential (\citealt{pearson15}, \citealt{price16}, Yavetz et al., {\it in prep.}).

The fact that useful information can be extracted from the morphology of thin GC streams alone opens up the exciting possibility of applying some of the intuition built for streams around our own Milky Way to other galaxies.  GC streams, however, have lower masses, are thinner than streams from dwarf galaxies, and are therefore harder to detect in external galaxies. Interestingly, \citet{abraham18} reported a detection of the ``Maybe Stream" with the Hubble Space Telescope (HST), which they suggest could be a GC stream 20 Mpc away. 

In this paper, we investigate current and future prospect of observing thin, globular cluster streams in external galaxies both through resolved stars and integrated light. 
Specifically, we ask whether globular cluster streams will be observable in resolved stars with upcoming telescopes such as WFIRST (\citealt{spergel13}, \citealt{spergel15}), or in integrated light with current telescopes such as the Hyper Suprime-Cam (HSC: \citealt{miyazaki12}) and future telescopes such as LSST (\citealt{ivezi08}) and Euclid (\citealt{racca16}). 

The paper is organized as follows: in Section \ref{sec:coldstreams}, we describe the properties of globular cluster streams and how we create mock-streams to test whether they are observable. In particular we describe the Palomar 5 (Pal 5) globular cluster stellar stream which we use as our fiducial model (Section \ref{sec:pal5}), we describe how we populate our streams with stars (Section \ref{sec:lum}), and how we calculate widths and lengths of mock-streams at various galactocentric radii (Section \ref{sec:length}). In Section \ref{sec:results}, we present the results on detecting streams in resolved stars in the Andromeda galaxy (M31) (Section \ref{sec:resolved}), in other external galaxies (Section \ref{sec:resother}) and in integrated light (Section \ref{sec:integrated}). We discuss the implications of our results in Section \ref{sec:discussion} and conclude in Section \ref{sec:conclusion}.

\section{Cold globular cluster stellar streams}
\label{sec:coldstreams}
Our goal is to estimate the observability of GC stellar streams in external galaxies. In this Section, we describe the framework we use to create mock stellar streams.  We chose the Milky Way's stellar stream, Pal 5,  as our reference stream, because the progenitor cluster of Pal 5's stream has been observed, which is only the case for very few known streams. Knowing the progenitor, enables us to determine the properties of the overall system more precisely (such as age, mass, metallicity, and orbit).

\begin{figure*}
\centerline{\includegraphics[width=\textwidth]{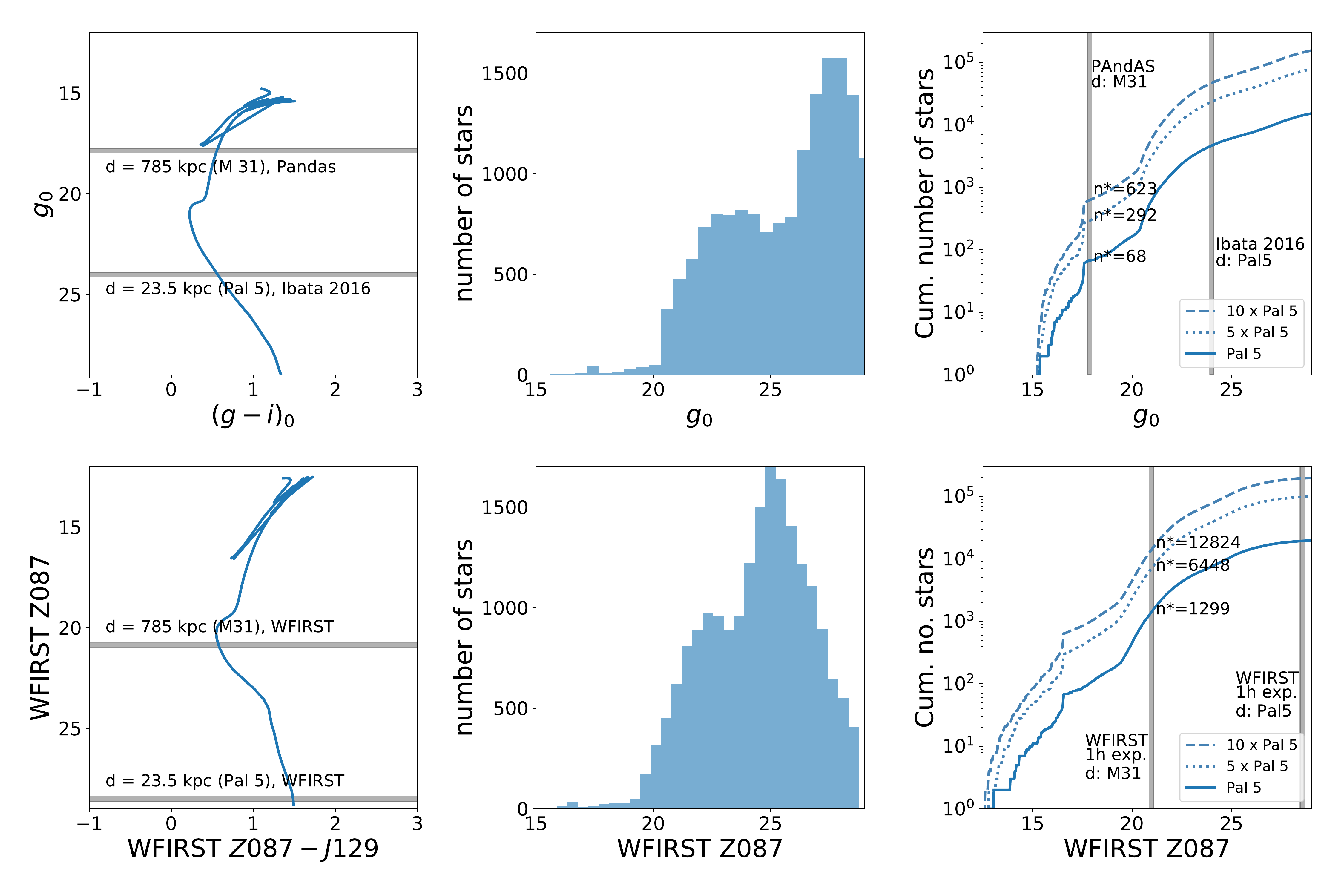}}
\caption{
{\bf Top Left:} CHFT $(g, i)_0$ color-magnitude diagram (CMD) showing a Pal 5-like cluster isochrone from the PARSEC system moved to the distance of Pal 5 in the Milky Way ($d$ = 23.5 kpc, $d_{mod} = 16.86$: \citealt{dotter11}). We use the same values for the cluster as \citet{ibata17}: Age = 11.5 Gyr and [Fe/H] $= -1.3$. The two horizontal lines show the faintest magnitude CHFT can observe ($g_0 < 24.0$) at Pal 5's current distance in the Galaxy, and the faintest magnitude PAndAS can observe ($g_0 < 25.5$) in M31 ($d = 785$ kpc, $d_{mod} = 24.47$). Note that at M31's distance, only the brightest red giant branch stars can be observed. 
{\bf Top Middle}: The luminosity function for the PARSEC Pal 5-like cluster normalized such that the amount of stars between $20 < g_0 < 23 = 3000$ (Bonaca et al., {\it in prep.}). 
{\bf Top Right:} The cumulative number of stars in a Pal 5-like stream (solid line), a 5 $\times$  more massive Pal 5-like stream (dotted line), and a 10 $\times$  more massive Pal 5-like stream (dashed line) for a given limiting $g_0$-mag. The vertical lines show the limiting magnitude of PAndAS ($g_0 < 25.5$) at the distance of M31 (i.e. shifted by 7.66 magnitudes from Pal 5's current location), and the limiting magnitudes of CHFT ($g_0 < 24$) at the distance of Pal 5 in the Milky Way. We indicate the amount of stars PAndAS should be able to observe for the Pal 5-like (n$^*$ = 68 $\pm~ 8$), 5  $\times$  Pal 5-like (n$^*$ = 292 $\pm~ 43$) and 10 $\times$ Pal 5-like stream (n$^*$ = 623 $\pm~ 84$) in M31 (see n$^*$). 
{\bf Bottom Left:} This row shows the same as the top row but for a Pal 5-like isochrone from WFIRST bands ($Z087$ and $J129$) and WFIRST limiting magnitudes for a 1 hour exposure ($Z087 <$ 28.54). {\bf Bottom middle:} To compute the WFIRST luminosity function, we sampled the exact same initial masses from the WFIRST isochrone downloaded from the PARSEC system as for the $g_0$-band above. 
Therefore, the WFIRST luminosity function is normalized to the Pal 5 stream between $20 < g_0 < 23$. 
{\bf Bottom right:} For an exposure time of 1 hour,
WFIRST will be able to observe n$^*$ = 1299 $\pm~ 93$ stream stars at the distance of M31 for a Pal 5-like cluster, n$^*$=  6448 $\pm~ 486$ stream stars for the 5 $\times$  more massive cluster, and n$^*$ = 12824 $\pm~ 966$ stream stars for a stream emerging from a cluster with 10 $\times$  the mass of Pal 5. All of these stars should be near the turn-off, on the red giant branch and later evolutionary stages (see left panel).}
\label{fig:iso_cfht}
\end{figure*}

\subsection{Palomar 5 data}
\label{sec:pal5}
 \citet{ibata16} presented photometric data of Pal 5 taken with the MegaCam instrument at the 3.6m Canada-France-Hawaii Telescope (CFHT) during 2006-2008. The CFHT $g, r$ bands provide data down to $g_0$ = 24 with good precision around the cluster main-sequence turnoff ($g_0$ refers to the extinction corrected value for $g$). In particular, their sample of stars with 20 $< g_0 <$ 23 has a completeness of 80\% (see \citealt{ibata16} and \citealt{ibata17} for more details). From \citet{ibata16} figure 7, we estimate that there are $\sim$1800 stars in the stream between 20 $< g_0 <$ 23 over a length of 20 $\deg$  and a width of 0.14 $\deg$ if we exclude the cluster stars.

Recently, Bonaca et al. ({\it in prep.}) analyzed Pal 5 in the photometric catalog of the DECam Legacy Survey (DECaLS) which is a part of the DESI Legacy Imaging Surveys (\citealt{dey2019}). They found that there were $3000 \pm 150$ stars in the Pal 5 stream between 20 $< g_0 <$ 23 excluding the cluster and after subtracting the background. Both Bonaca et al. ({\it in prep.}) and  \citet{ibata16} reach a similar limiting magnitude of ($g_0 < 24$), and the discrepancy in the number of stars  between 20 $< g_0 <$ 23 for the two data sets is likely due to a better star/galaxy separation in DECaLS. Throughout the paper, we use $3000 \pm 150$ stars between 20 $< g_0 <$ 23 to estimate how many stream stars there are in Pal 5's stellar stream at any given magnitude (see section below).

\subsection{Isochrones \& luminosity functions of streams}
\label{sec:lum}
To construct mock-streams, we first download isochrones for a Pal 5-like cluster from the PARSEC evolutionary tracks (\citealt{bressan12}, \citealt{margio17}). We use the same values for the Pal 5-like cluster as \citet{ibata17}: the age of the cluster is set to 11.5 Gyr and the metallicity is fixed as [Fe/H] $= -1.3$. We download the isochrone for both CFHT bands and WFIRST bands and shift the isochrones to Pal 5's current distance in the Milky Way ($d$ = 23.5 kpc, $d_{mod} = 16.86$: \citealt{dotter11}). We plot the shifted isochrone in the CFHT $g_0$ and $i_0$-bands in the upper left panel of Figure \ref{fig:iso_cfht}.  We also plot horizontal lines indicating which part of the isochrone is observable (the part of the isochrone that is above the vertical lines) given CFHT's limiting magnitude  ($g_0 < 24$) at Pal 5's location in the Milky Way, and given PAndAS' limiting magnitude ($g_0 < 25.5$) in M31 (see the top horizontal line). 

Subsequently, we compute a luminosity function, by sampling masses ($m = 0.01 - 120 ~\msun$) from a power law initial mass function (IMF: $dN/dm \propto m^{-0.5}$ as used by \citealt{ibata16}), and use the isochrone tables for age = 11.5 Gyr and [Fe/H] $= -1.3$ to infer the stars' magnitudes. 

To determine how many stars we can observe in a Pal 5-like stream at a given limiting magnitude, we normalize the luminosity function from CFHT such that there are  $3000$ stars (Bonaca et al., {\it in prep.}) with $20 < g_0 < 23$ (see Figure \ref{fig:iso_cfht}, top, middle panel). We then compute the cumulative number of stars at a given limiting magnitude (see Figure \ref{fig:iso_cfht}, top right panel, solid line).

Globular clusters have a spread in masses (\citealt{harris96}), and many are more massive than the inferred initial mass of Pal 5 (${\sim} 50,000 ~\msun$: \citealt{ibata17}). We therefore repeat the above exercise for streams with five (dotted line) and ten (dashed line) times the current stellar mass in the Pal 5 stream, where we resample the IMF such that there are 15,000 stars between $20 < g _0< 23$  (five times more than for the observed Pal 5-stream) and 30,000  stars between $20 < g_0 < 23$ (ten times more than for the observed Pal 5-stream). Note that even more massive globular cluster streams exist (see e.g. \citealt{ibata19b}), and that our estimates are therefore conservative.

In the top, right panel of Figure \ref{fig:iso_cfht}, we show that the PAndAS survey (\citealt{McConnachie18}, see more details in Section \ref{sec:PANDAS}) should detect $68 \pm 8$ stars in a Pal 5-like stream at the distance of M31 ($d_{mod} = 24.46$, see vertical line), $292 \pm$ 43 stars for a stream that is five times more massive than Pal 5, and $623 \pm$ 84 stars for a stream that is ten times more massive than Pal 5. The scatter arises due to the IMF sampler, and we have sampled the three IMFs 100 times to estimate the 1$\sigma$ deviation from the mean value. All of the observable stars are at brighter magnitudes than the main sequence turnoff (see left panel of Figure \ref{fig:iso_cfht}) and are in the red giant branch (RGB) or more evolved stellar evolutionary stages. The two vertical lines show the limiting magnitude of the PAndAS  survey if Pal 5 were residing in M31, and the limiting magnitude of \citet{ibata16}'s CHFT Pal 5 survey at the distance of Pal 5's current location in the Milky Way. 

In the bottom row of Figure \ref{fig:iso_cfht}, we show the same panels as in the top row, but now using downloaded PARSEC Pal 5-like isochrones in WFIRST $Z087$ and $J129$-bands. We used the same properties of the Pal 5-like cluster as described above (i.e. Age = 11.5 Gyr and [Fe/H] $= -1.3$). 

To estimate how many stars WFIRST should be able to observe in a Pal 5-like stream, we normalized the luminosity function by sampling the exact same initial masses in the dowloaded WFIRST isochrone as for the $g_0$-band CFHT masses. This provides us with the number of stars WFIRST will observe at a given magnitude. Throughout the paper, we use a 1 hour exposure time to determine the limiting magnitude of WFIRST.  For the $Z087$-band this is $Z087 <$ 28.54. In {\it Appendix} \ref{sec:appendixA}, we show how our results differ if we instead use the WFIRST guest observer capabilities of a 1000 second exposure, where the limiting $Z087$-band magnitude is  $Z087 <$ 27.15 (\citealt{spergel13}). 

At the distance of M31 and with a 1 hour exposure time, WFIRST will be able to detect $1299\pm 93$ stars in a Pal 5-like stream,  6448 $\pm~ 486$ stars for a stream that is five times more massive than Pal 5, and $12824 \pm 966$ stars for a stream that is ten times more massive than Pal 5 (Figure \ref{fig:iso_cfht}, bottom right panel). 
We explore the observability of resolved stars in GC streams in M31 with the PAndAS survey and with WFIRST in Section \ref{sec:resolved}. 
Note that the tip of the RGB in the $Z087$-band for a Pal 5-like stream is at an absolute magnitude of $-4.3$. Thus, at distances $\geq$38 Mpc ($d_{mod} \geq$ 32.9), even the brightest giants will not be resolved in the $Z087$-band for a 1 hour exposure with WFIRST, as the limiting magnitude is $Z087 <$  28.54 (see Figure \ref{fig:iso_cfht}, lower left).

\begin{table*}
\centering
\caption{Properties of mock-streams in an M31-like halo\footnote{$n_*$ denotes the number of stars, $l$: the length, $w$: the width, $Area$: the area, $\mu$: the surface brightness of each mock-stream.}}
\label{tab:Pal5}
\begin{tabular}{lccc}
\hline
 & {\bf R$_{GC}$ = 15 kpc }&  {\bf R$_{GC}$ = 35 kpc} &  {\bf R$_{GC}$ = 55 kpc} \\ 
 \hline
{\bf Pal 5-like mass} & &&\\
$n_*$ (1000 sec.,  $Z087 <$ 27.15) & 259 $\pm 17$ & \\
$n_*$ (1 hour, $Z087 <$ 28.54) &1299 $\pm$ 93& \\
$l$ [kpc] & 7.81 & 10.4 &  12.0  \\
$w$ [kpc] & 0.053 & 0.094  & 0.127  \\
$Area$ [kpc$^2$] & 0.414 & 0.973 &1.53 \\
$\mu$ [mag arcsec$^{-2}$] ($R062$-band) &32.7 $\pm$ 0.12 &33.7 $\pm$ 0.12 &34.1 $\pm$ 0.12\\
$\mu$ [mag arcsec$^{-2}$] ($Z087$-band) &32.0 $\pm$ 0.16 &33.0 $\pm$ 0.16 &33.4 $\pm$ 0.16 \\
$\mu$ [mag arcsec$^{-2}$] ($F184$-band) &30.8 $\pm$ 0.23 &31.8 $\pm$ 0.23&32.3 $\pm$ 0.23\\

\hline
{\bf 5 $\times$ Pal 5-like mass} & &&\\ 
$n_*$ (1000 sec., $Z087 <$ 27.15) &1233 $\pm 86$ & \\
$n_*$ (1 hour, $Z087 <$ 28.54) &6448 $\pm$ 486& \\
$l$ [kpc] &13.3 & 17.7&20.6  \\
$w$ [kpc]& 0.091 &  0.16& 0.217  \\
$Area$ [kpc$^2$] & 1.2  &  2.8  &  4.5\\
$\mu$ [mag arcsec$^{-2}$] ($R062$-band) & 32.2 $\pm$ 0.06 &33.1 $\pm$ 0.06 &33.6 $\pm$ 0.06\\
$\mu$ [mag arcsec$^{-2}$] ($Z087$-band) & 31.4 $\pm$ 0.08 &  32.4 $\pm$ 0.08& 32.9 $\pm$ 0.08 \\
$\mu$ [mag arcsec$^{-2}$] ($F184$-band) &30.2 $\pm$ 0.11 &31.2 $\pm$ 0.11 &31.7 $\pm$ 0.11 \\

\hline
{\bf 10 $\times$ Pal 5-like mass} & &&\\ 
$n_*$ (1000 sec., $Z087$ $<$ 27.15) &2440 $\pm 165$ & \\
$n_*$ (1 hour, $Z087$ $<$ 28.54) &12824 $\pm$ 966& \\
$l$ [kpc] &16.8 & 22.3 & 25.9 \\
$w$ [kpc]& 0.115 & 0.202 & 0.273  \\
$Area$ [kpc$^2$] &  1.93& 4.51   & 7.08  \\
$\mu$ [mag arcsec$^{-2}$] ($R062$-band) & 31.9 $\pm$ 0.04 &32.8 $\pm$ 0.04 &33.3 $\pm$ 0.04\\
$\mu$ [mag arcsec$^{-2}$] ($Z087$-band) &31.2 $\pm$ 0.05 &32.1 $\pm$ 0.05 &32.6 $\pm$ 0.05 \\
$\mu$ [mag arcsec$^{-2}$] ($F184$-band) & 30.0 $\pm$ 0.07& 30.9 $\pm$ 0.07 & 31.4 $\pm$ 0.07\\

\hline 
\end{tabular}
\end{table*}

\subsection{Length \& width of streams}
\label{sec:length}
In this Section, we describe how we populate mock-streams with the given number of stars found in the previous section. In particular, we compute the widths and lengths of mock GC streams, taking Pal 5 as our starting point. 

To place a Pal 5-like stream in the M31 halo, we scale mock-streams from the initial length and width of the Pal 5 stream in our Galaxy: length = 8.5 kpc, width = 58 pc (\citealt{ibata16}). For simplicity, we assume a circular orbit with a distance from the center of the Galaxy of R$_{GC}$ = 15 kpc, which is similar to the average of Pal 5's apocentric and pericentric distance for its rather elliptical orbit (see e.g. \citealt{erkal17}). 

We follow \citet{johnston98} and \citet{johnston01}, to compute the width and lengths of the streams at three different galactocentric radii: $R_{GC}$ = 15 , 35, 55 kpc in M31's halo. We compute lengths and widths for all the streams with our different stream masses: 1) a Pal 5-like stream mass 2) a five times more massive Pal 5-like stream 3) a ten times more massive Pal 5-like stream. Note that we do not  assume a mass for the cluster, but scale stream properties based on the observed properties of the Pal 5 stream in the Milky Way today (i.e. the length and width presented in \citet{ibata16} and the number of stream stars presented in Bonaca et al., {\it in prep.}, excluding the stars in the cluster itself).  

Following \citet{johnston01} equation 8, we can express the width, $w$, of the streams as:

\begin{equation}
w = R_{GC} \left[\frac{m}{M(R_{GC})}\right]^{1/3} = R_{GC} \left[\frac{m G}{v_c^2 R_{GC}}\right]^{1/3} 
\end{equation}
where $R_{GC}$ is the radius of the orbit of the globular cluster around its host (which is normally expressed as $R_p$, but since we have a circular orbit, this remains constant over the entire orbit), $m$ is the mass of the cluster, $M(R_{GC})$ is the enclosed mass of the host within the stream's orbit, $v_c$ is the circular velocity at radius $R_{GC}$, and $G$ is the gravitational constant. We assume a flat rotation curve and therefore constant  $v_c$, which is a valid assumption at $R_{GC} > 15$ kpc in M31 (e.g. \citealt{chemin09}). Thus the width of a GC stream scales as:

\begin{equation}
\label{eq:w}
w \propto R_{GC}^{2/3} \frac{m^{1/3}}{v_c^{2/3}}
\end{equation}

We scale the width of Pal 5 based on each parameter in Equation \ref{eq:w} separately. First, we update the width of the stream reflecting the fact that the circular velocity in M31 is $v_{c, M31}= 250 ~\kms$ at $R_{GC} = 15$ kpc (\citealt{chemin09}) as opposed to $v_{c, MW}= 217 ~\kms$ for the Milky Way (\citealt{eilers19}). Hence, we correct the width to be narrower by a factor of $v_{corr} =  \left(\frac{v_{c,M31}}{v_{c,MW}}\right)^{-2/3} = 0.91$.

Subsequently, we update the width of the stream based on the location in M31's halo. Thus, as the stream is placed at larger radii, we enlarge the width by a factor of  $R_{corr} = \left(\frac{R_{GC}}{R_{GC,Pal5}}\right)^{2/3}$. Lastly, we scale the mass up by a factor of five and ten and therefore make the stream a factor of $m_{corr} = \left(\frac{5m}{m}\right)^{1/3}= 1.71$  and $m_{corr} = \left(\frac{10m}{m}\right)^{1/3}= 2.15$ wider for the more massive streams. Recall that we do not assume anything about the Pal 5 cluster's initial or present day mass. 

In this work, we do not include the fact that the stream should be wider towards its endpoints. Instead, we sample the stream stars from a normal distribution, with a dispersion (width) calculated using Equation \ref{eq:w} and listed in Table \ref{tab:Pal5} (see also \citealt{ibata16}, fig. 12).

Following \citet{johnston01} equation 5, we can express the angular length, $\Psi$, of the stream as:
\begin{equation}
\Psi = 4  \left[\frac{m}{M(R_{GC})}\right]^{1/3}  \frac{2 \pi t}{T_{\psi}} \propto \left[\frac{m}{M(R_{GC})}\right]^{1/3}  \frac{t}{T_{\psi}}
\end{equation}
where $t$ is the dynamical age of the stream, and $T_{\psi}$ is the azimuthal period of the cluster around its host galaxy. We can re-write the angular length as:
\begin{equation}
\Psi \propto \left[\frac{m }{v_c^2 R_{GC}}\right]^{1/3}  \frac{v_c t }{R_{GC}} = \frac{m^{1/3}}{R_{GC}^{4/3}} v_c^{1/3}t
\end{equation}
The physical length of the stream can thus be expressed as:
\begin{equation}
\label{eq:l}
L \propto R_{GC} \frac{m^{1/3}}{R_{GC}^{4/3}} v_c^{1/3}t = \frac{m^{1/3}}{R_{GC}^{1/3}} v_c^{1/3}t
\end{equation}

We scale the length of Pal 5 based on each parameter in Equation \ref{eq:l} separately. Note that we assume the same age of the stream in all cases (t = constant). We first make the stream longer by a factor of $(\frac{v_{c,M31}}{v_{c,MW}})^{1/3} = 1.05$. Subsequently, we update the length of the stream based on the location in M31's halo. Thus, as the stream is placed at larger radii, we shorten the length by a factor of  $R_{corr} = \left(\frac{R_{GC}}{R_{GC,Pal5}}\right)^{-1/3}$, reflecting the fact that these streams will have completed less orbits around their hosts.
Additionally, we scale the length of the stream by a factor of $m_{corr} = \left(\frac{5m}{m}\right)^{1/3}$ and $m_{corr} = \left(\frac{10m}{m}\right)^{1/3}$ for the more massive streams. 

To summarize, we now have the framework to construct mock stellar streams of various lengths and widths scaled from Pal 5's initial stream properties. We populate nine different mock-streams of various widths and lengths on great circles with radii of $R_{GC} =$ 15, 35, 55 kpc, and with the number of stars obtained in Section \ref{sec:lum} (see Table \ref{tab:Pal5}). Note that there are several ways we could have constructed mock-streams, and that the mock-streams presented in this work are meant to illustrate a range of possible streams. In reality, the stellar streams could have a slightly different range of orbital properties, lengths, widths and number of resolved stars depending on the initial cluster's location, mass, age and metallicity.

\begin{figure*}
\centerline{\includegraphics[width=\textwidth]{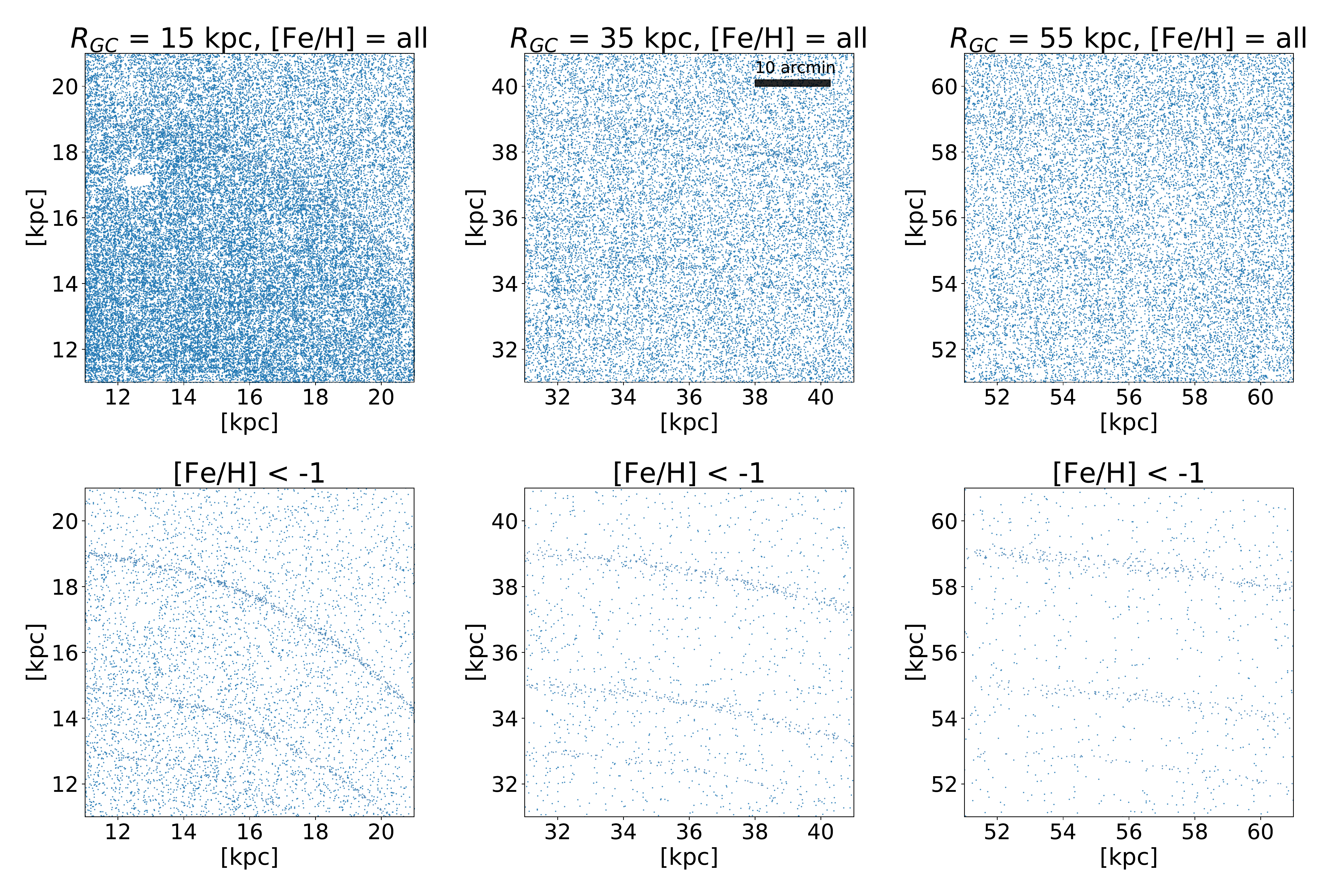}}
\caption{
In this plot we demonstrate the observability of a Pal 5-like stream, a 5 $\times$  more massive Pal 5-like stream, and a 10 $\times$  more massive Pal 5-like stream for the PAndAS survey of the M31 halo. Each column shows the PAndAS data at different M31 galactocentric radii (left: $R_{GC}$ = 15  kpc, middle: $R_{GC}$ = 35 kpc, right: $R_{GC}$ = 55 kpc). The fields are shown in kpc and are converted from an angular scale assuming a dustance of  785 kpc to M31. The black bar in the top middle panel shows the scale of 10 arcmin for reference. 
In the first row we show all stars observed in the PAndAS fields, and in the second row we show all stars observed in PAndAS with [Fe/H] $< -1$. In each panel, we have injected three streams: a Pal 5-like stream, a Pal 5-like stream with 5 $\times$ the mass of Pal 5, and a Pal 5-like stream with 10 $\times$  the mass of Pal 5. Recall that the Pal 5-like stream has [Fe/H] $= -1.3$, and should therefore be visible in both rows.  
We determine the number of stars in the three mock-streams by summing up the cumulative number of stars in the streams at the limiting magnitude of the PAndAS survey ($g_0 < 25.5$) at the distance of M31 (see Figure \ref{fig:iso_cfht}, right panel). There are $623$ stars in the ten times more massive stream,  $292$ stars in the five times more massive stream and only $68$ stars in Pal 5-like stream. At each $R_{GC}$, we have updated the width and lengths of the mock-streams based on the tidal field they experience at these distances and based on their masses (see Section \ref{sec:length}). We note that hints of the Pal 5-like stream is visible in the lower panels, and that the 5 and 10 $\times$  more massive stream becomes apparent in the halo when we apply a metallicity cut (see middle and right panels in the second row).}
\label{fig:M31_pandas}
\end{figure*}

\section{Results}\label{sec:results}
In this Section, we present results on the detectability of GC stellar streams for resolved stars in M31 (Section \ref{sec:resolved}), for resolved stars in other external galaxies (Section \ref{sec:resother}), and for integrated light (Section \ref{sec:integrated}).

\subsection{Cold streams in M31: resolved stars}
\label{sec:resolved}
We first explore whether GC streams should stand out against the background stellar halo of M31 given the PAndAS limiting magnitudes (Section \ref{sec:PANDAS}), and we test the same scenario given WFIRST's limiting magnitudes and bands (Section \ref{sec:WFIRST}).

\subsubsection{PAndAS}
\label{sec:PANDAS}
The PAndAS survey (e.g. \citealt{McConnachie18}) has mapped a total area of 400 square degrees surrounding M31 using the 1-square-degree field-of-view MegaPrime/MegaCam camera on the 3.6m Canada-France-Hawaii Telescope. It surveyed in $g, i$-bands to depths of $g$ = 26.5, $i$ = 25.5, resolving individual stars with a signal-to-noise ratio of 10. The PAndAS team derive photometric metallicities for all stars by assuming that the width of the red giant branch (RGB) can be interpreted as the spread in metallicity within a galaxy (see e.g. \citealt{crno14}).

The columns in Figure \ref{fig:M31_pandas} show 10 $\times$ 10 kpc regions from the PAndAS survey of M31's halo at various galactocentric radii ($R_{GC}$ = 15  kpc, $R_{GC}$ = 35 kpc and $R_{GC}$ = 55 kpc). All plotted stars have $g_0 < 25.5$, and the two rows show the same areas without a metallicity cut: [Fe/H] = all (top row) and after we have applied a metallicity cut of  [Fe/H] = $< -1 $ (bottom row). Recall that the Pal 5 cluster metallicity is  [Fe/H] $= -1.3$, and that all Pal 5 stars will therefore be visible for these metallicity cuts. Note also that PAndAS has and WFIRST will have photometrically determined metallicities, so a metallicity cut is in principle equivalent to a color and magnitude cut.

In each panel, we have injected a Pal 5 mock stellar stream and two mock stellar streams with properties scaled to have 5 and 10 times  the mass of Pal 5. We have populated the streams with stars using the PAndAS limiting magnitude calculated in Section \ref{sec:lum} (68, 292 and 623 stars).  
The streams' lengths and widths have been calculated based on the streams' distances from M31's galactic center and the streams' masses (see Section \ref{sec:length}). 

In Figure \ref{fig:M31_pandas}, we see hints of the Pal 5-like stream and the 5 times more massive stream in the lower panels of Figure \ref{fig:M31_pandas} after we have applied metallicity cuts. The 10 times more massive stream is visible by eye in each panel, but hard to see at large galactocentric radii without metallicity cuts as the stream is wider (see upper right panel)\footnote{Note that we have located the mock-streams at the same positions in both Figure \ref{fig:M31_pandas} and \ref{fig:M31_wfirst}, in case the reader is curious to know where the Pal 5-like streams are located in Figure \ref{fig:M31_pandas}.}.

The PAndAS team has not yet reported a detection of massive GC streams, thus M31might not host GC streams that are much more massive than Pal 5's stream. It should be feasible, however, to find a 5 and 10 times more massive Pal 5-like stream in the PAndAS data, should such a stream exist in M31. Using more sophisticated searching methods and cuts, a Pal5-like stellar stream may also be detectable (see the discussion in Section \ref{sec:color}).

\subsubsection{WFIRST}
\label{sec:WFIRST}
\begin{figure*}
\centerline{\includegraphics[width=\textwidth]{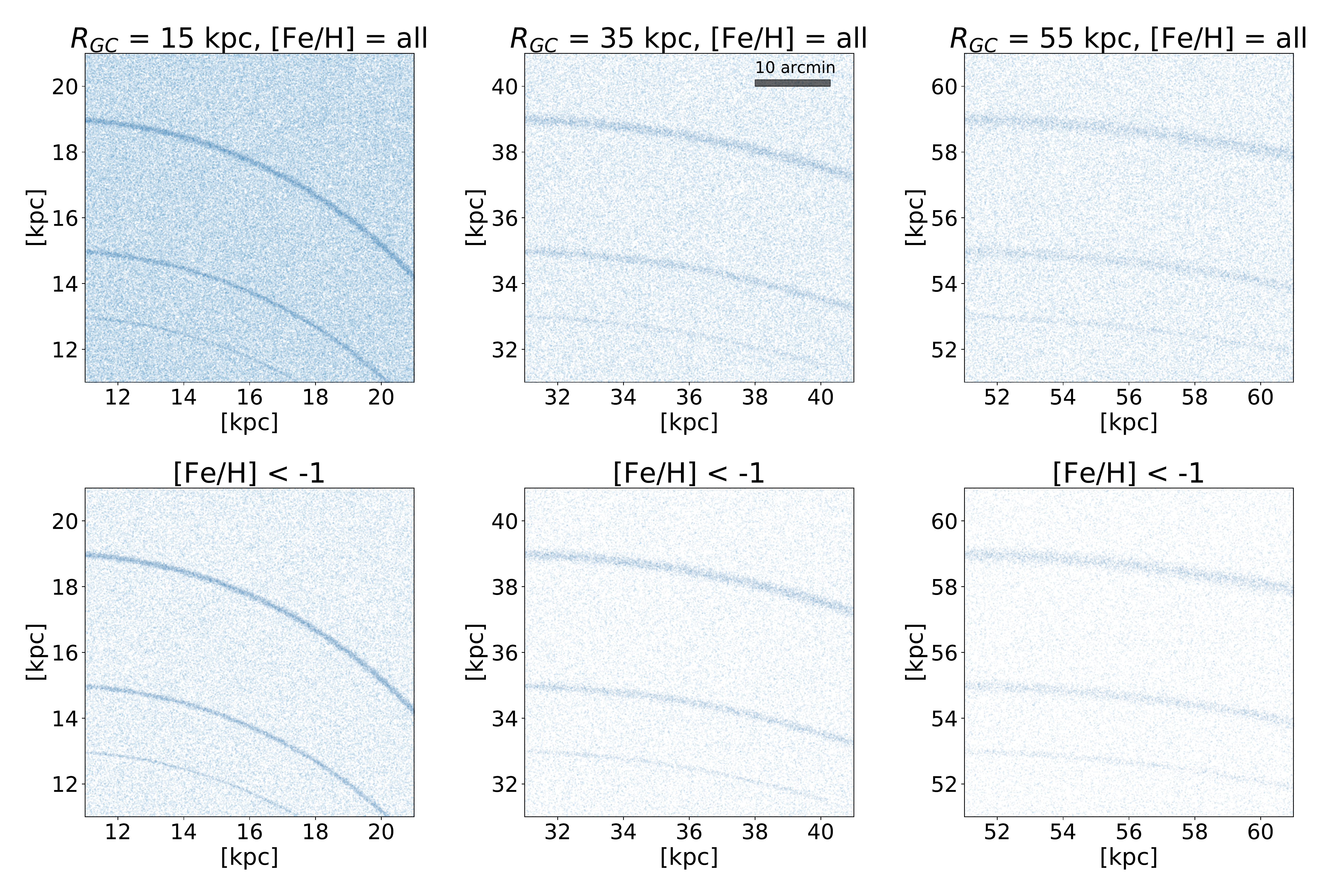}}
\caption{
Like Figure \ref{fig:M31_pandas}, but we now determine the number of stars in each mock-stream by summing up the cumulative number of stars in the streams at the limiting magnitude of WFIRST ($Z087 < 28.54$) at the distance of M31. Given the large number of stars in the background fields and streams, stars are overlapping in these panels, and we use a point size 100 times smaller than in Figure \ref{fig:M31_pandas} to avoid crowding of the background. WFIRST will resolve each individual stream star (see Section \ref{sec:crowding}), and the point sizes in this Figure are larger than the WFIRST pixel sizes of 0.11$\arcsec \times 0.11\arcsec$. Thus, visualizing the stream stars as overlapping points, as the case in this Figure, will be feasible with WFIRST data.
The black bar in the top middle panel shows the scale of 10 arcmin for reference. At each $R_{GC}$, we have updated the width and lengths of the streams based on the tidal field they experience at these distances (see Section \ref{sec:length}). Additionally, we have updated the number of stars in each field such to illustrate what WFIRST will observe in M31 given WFIRST's  deeper limiting magnitude (see details in Section \ref{sec:WFIRST}). In particular, the ten times more massive stream is populated with 12824 stars, the five times more massive stream has 6448 stars, and the Pal 5-like stream has 1299 stars (see Figure \ref{fig:iso_cfht}, bottom, right panel).  All stream stars have [Fe/H] $= -1.3$, and are therefore plotted in each panel. 
We note that all mass mock-streams are observable in each panel.}
\label{fig:M31_wfirst}
\end{figure*}
NASA's Wide Field InfraRed Survey Telescope (WFIRST) is planned to launch in mid-2020s. The space telescope has a 2.4 meter primary mirror and its wide field instrument will have a field of view that is a hundred times greater than the Hubble infrared instrument and has a pixel size of 0.11$\arcsec$. Thus, WFIRST is particularly useful for studying resolved stellar populations over large areas. 

To test whether a Pal 5-like stream will be detectable with WFIRST, we first simulate M31's halo for WFIRST bands and limiting magnitudes. We do this by populating the 10 $\times$ 10 kpc PAndAS panels from Figure \ref{fig:M31_pandas} ($R_{GC}$ = 15, 35, 55 kpc) with additional stars to mimic the densities and magnitudes WFIRST will observe. To do this, we first download a grid of isochrones, 16 in total, from the PARSEC system with a metallicity range of [Fe/H] = $-2.5$ - 0.5 in increments of [Fe/H] = 1 dex, and an age range of age = 8.5 - 11.5 Gyr in increments of 1 Gyr. Subsequently, we fit all stars in the color-magnitude diagrams of the three 10 $\times$ 10 kpc  PAndAS M31 panels to these isochrones at the distance of M31, and at a range of distances representative of the Milky Way disk and halo from ${\sim}500$~pc to ${\sim}100$~kpc to fit foreground stars. Using the number of stars fitted to each isochrone and the luminosity functions of the 16 isochrones, we then estimate  for each individual isochrone the number of stars needed to match the stellar population visible for a 1 hour exposure with WFIRST ($Z087 <$ 28.54).

Our mock 10 $\times$ 10 kpc WFIRST M31 halo fields are then populated with these stars with random positions within the field, ages and metallicities corresponding to their respective isochrone, and magnitudes sampled from the part of the luminosity function of the isochrone that is observable within WFIRST limits.

Additionally, we convert the deep optical fields from the \citet{brown09} M31 HST halo fields (F606W $<$ 32) to WFIRST bands and apply the limiting magnitude of the WFIRST. We then use the stellar density in the three \citet{brown09} halo fields, which are located at $R_{GC}$ = 11, 21, 35 kpc to estimate the stellar density that WFIRST will observe at the location of the 10 $\times$ 10 kpc panels ($R_{GC}$ = 15, 35, 55 kpc), using a power-law fit to the M31 stellar density profile. We use these stellar densities to check the stellar densities for our new M31 halo fields for WFIRST limits. From the power-law fit to the Brown fields stellar densities, the stellar densities that WFIRST will observe in the M31 halo, are $\rho_*  \sim 7 \times 10^5$,  ${\sim}7 \times 10^4$,  ${\sim} 2 \times 10^4$ stars degrees$^{-2}$ at  $R_{GC}$ = 15, 35, 55 kpc, respectively. The stellar densities that we reach at the same radii by fitting isochrones and adding fake stars are  $1.2 \times 10^5$, $5.1 \times 10^4$, and $4.5 \times 10^4$ stars degrees$^{-2}$ at  $R_{GC}$ = 15, 35, 55 kpc, respectively. Thus, the fact that the stellar densities in the \citet{brown09} fields are similar to the stellar densities in our simulated fields, after applying WFIRST's magnitude cuts, is encouraging. 

Figure \ref{fig:M31_wfirst} shows the same panels as Figure \ref{fig:M31_pandas}, however the background fields now represent WFIRST limiting magnitudes and stellar densities. We inject the same mock stellar streams as in Figure \ref{fig:M31_pandas}, but now we populate the streams with a larger amount of stars reflecting the limiting magnitudes and bands of WFIRST (see bottom row of Figure \ref{fig:iso_cfht}). In particular, we populate the Pal 5-like stream with 1299 stars, the five times more massive stream with 6448 stars, and the ten times more massive stream with 12824 stars.

All mock-streams are visible in all panels of Figure \ref{fig:M31_wfirst}. This provides exciting prospects for the detection of thin, stellar streams in M31's halo and could vastly expand our sample of known GC streams (see Section \ref{sec:kara}) for which we can investigate the morphology and search for gaps in streams. We note that the results of  Figure \ref{fig:M31_wfirst} are based on the evolutionary tracks of an old globular cluster (Pal 5-like) and are based on a 1 hour exposure time for WFIRST. We refer the reader to {\it Appendix} \ref{sec:appendixA} for details on how a 1000 sec. exposure, which is the guest observer capability of WFIRST (\citealt{spergel13}), would affect our results. 

A younger cluster than Pal 5, would add many more observable stars to the streams given WFIRST's limiting magnitudes, as more stars would have brighter magnitudes. As an example, we download an isochrone from the PARSEC evolutionary tracks (\citealt{bressan12}) for a cluster with age = 500 Myr and  [Fe/H] = 0. We repeat the exercise carried out to produce Figure \ref{fig:iso_cfht}. For this isochrone the number of resolved stars with WFIRST at M31's distance would be ${\sim}$7797 for a Pal 5-like stream and ${\sim}$38754 for a 5  times more massive stream and ${\sim}$77322 for a 10 times more massive stream. Thus, roughly a factor of 6 more stars. 
Note, however, that it would be very difficult to form a stream with a significant length 500 Myr after the birth of the cluster. 

\subsection{Cold globular cluster streams in external galaxies: resolved stars}
\label{sec:resother}
In addition to exploring WFIRST's capabilities for observing GC streams in M31, we wish to determine WFIRST's ability to observe GC streams in more distant, external galaxies. First, we investigate the impact of including contaminating galaxies for our ability to detect streams (Section \ref{sec:contamination}). Subsequently, we compute the threshold number of stars needed to detect streams in galaxies at various distances (Section \ref{sec:Cerr}). To test the validity of our results, we also investigate the impact of crowding in the WFIRST pixels (Section \ref{sec:crowding}).

\subsubsection{Contaminating galaxies}
\label{sec:contamination}
Our ability to detect faint GC streams with resolved stars will be significantly affected by the quality of our star/galaxy separation.  To assess how contaminating background galaxies will affect our ability to detect thin streams, we use the Space Telescope Image Product Simulator (STIPS\footnote{\url{https://github.com/spacetelescope/STScI-STIPS}}) to simulate one WFIRST detector which is $7.5\arcmin \times 7.5\arcmin$ in size centered on the $R_{GC} = 35$ kpc M31 field shown in the top, middle panel of Figure \ref{fig:M31_wfirst}, without an injected mock-stream.  While the limited field size available in STIPS is small compared to the stream and is not yet sufficient for full stream simulations, it is sufficient for testing completeness and contamination rates.  The simulations include an input foreground star-background galaxy catalog with a brightness distribution and a Sersic profile distribution generated from the CANDELS (\citealt{koekemoer11}) deep survey.  We then perform point spread function (PSF) fitting photometry on these images with the DOLPHOT software package, including the newly-developed WFIRST module (\citealt{dolphin00}; \citealt{dolphin16}).

Current techniques for culling non-point sources from resolved stellar photometry typically use cuts on quality metrics such as signal-to-noise, sharpness, and crowding (e.g. for the Panchromatic Hubble Andromeda Treasury; \citealt{williams14}).  Thus, to get a reasonable upper-limit estimate of the contamination we may expect from background galaxies using very simple quality metrics, we use a signal-to-noise cut $>$ 4, a sharpness$^2$ requirement $<$ 0.1, and a crowding limit $<$ 0.2 in the $Z087$ WFIRST band. We do this for both a 1 hour and 1000 second exposure time for WFIRST. 

The resulting stellar catalogs include the stars in the given M31 field, but also all of the objects remaining after our efforts of removing contaminants using the quality cuts listed above. In the coming years, we will be working to improve this culling by including color information and applying sophisticated machine learning techniques.  Such development will only improve the reliability of the star catalogs, reducing the leftover contaminants in this field.  Below, we explore the effects of using both perfect star/galaxy separation and ``worst case scenario" contamination from background galaxies (with simple catalog culling described above) when searching for GC streams in external galaxies in order to bracket the potential sensitivity of WFIRST to GC streams.
\begin{figure*}
\centerline{\includegraphics[width=\textwidth]{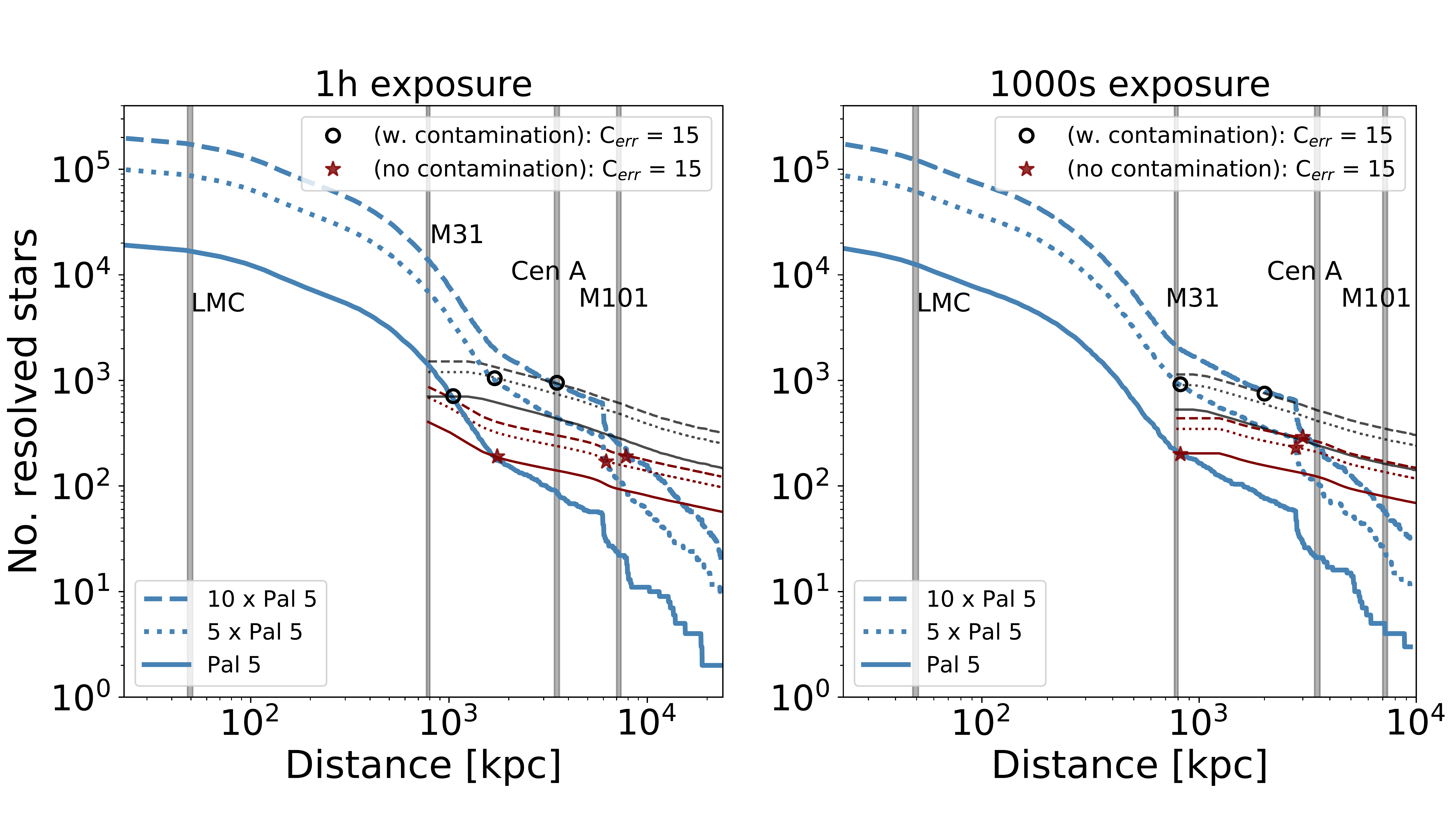}}
\caption{
{\bf Left}: The number of resolved WFIRST $Z087$-band stars in a Pal 5-like stream (solid blue line), a five times more massive Pal 5-like stream (dotted blue line) and a ten times more massive Pal 5-like stream (dashed blue line) as a function of distance with a 1 hour WFIRST exposure time ($Z087 < 28.54$). We calculate the number of resolved stars in each mock-stream at each distance from the cumulative luminosity function in the WFIRST $Z087$-band (see right, bottom panel of Figure \ref{fig:iso_cfht}). The vertical lines show the distances to four well-known galaxies within 10 Mpc of the Milky Way. 
The red lines shows the threshold number of stars needed in each of the three mass mock-streams, for the streams to have a signal-to-noise of  $C_{\rm err} = 15$, assuming perfect star/galaxy separation (no contamination). 
To compute the red lines for each mass mock-stream, we have placed each stream at a galactocentric radius of 35 kpc (to fix their areas), and we have used no metallicity for the calculations of $C_{\rm err}$. We mark the distances at which the red lines cross the blue lines for each respective mass mock-stream with a red star. 
The 5 and 10 times  Pal 5-like mock-streams have $C_{\rm err} > 15$ out to distances $\sim$6.2 and $\sim$7.8 Mpc respectively, and the Pal 5-like mock-stream has $C_{\rm err} > 15$ out to $\sim$1.8 Mpc.
The gray lines show the same calculation as for the red lines, but now using backgrounds including contaminating galaxies (see Section \ref{sec:contamination}). We mark the distances at which the gray lines cross the blue lines for each respective mass mock-stream with a black circle. In particular, a Pal 5- like stream will be observable out to $\sim$1.1 Mpc, and  5 and 10 times more massive streams will be observable out to  $\sim$1.7 and 3.5 Mpc, respectively.
Note that beyond 38 Mpc no stars will be resolved in the $Z087$-band for a 1 hour exposure with WFIRST in a Pal 5-like stream.
{\bf Right:} Same as left panel, but now using a 1000 sec. exposure time ($Z087 < 27.15$).  
Assuming prefect star/galaxy separation (red lines), the 5 and 10 $\times$  Pal 5-like mock-streams have $C_{\rm err} > 15$ out to distances $\sim$ 2.8 and 3.0 Mpc respectively, and the Pal 5-like mock-stream at $R_{GC} = 35$ kpc has $C_{\rm err} = 15$ at the distance of M31 (see also {\it Appendix} \ref{sec:appendixA}). If we include the effect of contaminating background galaxies, the Pal 5-like stream will not be easily detectable using a 1000 sec. WFIRST exposure at any distance as the gray solid line is above blue solid line at all distances. A simple extrapolation of the solid gray line, however, indicates that Pal 5-like streams could be detectable out to $\sim$400 kpc with a 1000 sec. exposure. 
The 5 and 10 times more massive streams will be observable out to $\sim$0.8 and 2.0 Mpc, respectively (see black circles). Note that beyond 20 Mpc, no stars will be resolved in the $Z087$-band in a Pal 5-like stream with the guest observer capabilities of WFIRST as even the RGB stars will have magnitudes fainter than $Z087$ = 27.15. The distance limit for detecting streams in external galaxies is likely somewhere in between the perfect star/galaxy separation scenario (red lines) and the conservative star/galaxy separation scenario (gray lines).
}
\label{fig:distance}
\end{figure*}

\subsubsection{Distance limitations for observing GC streams with WFIRST}
\label{sec:Cerr}
Only hints of the Pal 5-like stream were present in the PAndAS data (Figure \ref{fig:M31_pandas}). Motivated by this, we compute the signal-to-noise (i.e. the contribution of the stream signal to M31's background) at different galactocentric radii ($R_{GC} = 15, 35, 55$ kpc), with two different metallicity cuts ([Fe/H] = all and $< -1$).  In particular, we compute the counting error, $C_{\rm err}$: 

\begin{equation}
\label{eq:cerr}
{C}_{\rm err} = \frac{N_{\rm stream}}{\sqrt{N_{\rm background}}}
\end{equation}
where $N_{\rm stream}$ is the number of stars in the mock-streams, and $N_{\rm background}$ is the  averaged number of stars in the M31 background, covering the same area as the specific mock-stream (see Table \ref{tab:Pal5}).

In Figure \ref{fig:M31_pandas}, the Pal 5-like stream has $C_{\rm err} > 15$ in the three lower panels (in the lower right panel $C_{\rm err} \sim 15$), but has $C_{\rm err} <15$ in the top row. The 5 times more massive stream has a $C_{\rm err} > 10$ in the first row of Figure \ref{fig:M31_pandas}, while $C_{\rm err} > 15$ for all panels in the bottom row. For the 10 times more massive stream, $C_{\rm err} > 15$ in all six panels. For reference, in Figure \ref{fig:M31_wfirst}, all Pal 5-like mock-streams have $C_{\rm err} > 40$, all 5 times more massive streams have $C_{\rm err} > 100$, and all 10 times more massive streams have $C_{\rm err} > 170$.

Motivated by the fact that some streams become hard to see if $C_{\rm err} < 15$, we use $C_{\rm err} = 15$ as our threshold for determining when a stream is ``easily observable" with WFIRST. We note that the choice of $C_{\rm err} = 15$ is somewhat arbitrary and that observers estimating which streams they will be able to detect, can use a more or less conservative signal-to-noise threshold than $C_{\rm err} = 15$. 

Our goal is to estimate the distance to which WFIRST will be able to detect GC streams with  $C_{\rm err} > 15$. As we go to fainter magnitudes in WFIRST, contamination by galaxies will be an issue which is why spatial resolution is essential. We assess both the prospects for detecting streams in external galaxies assuming a perfect star/galaxy separation, and including the effects of contaminating background galaxies.

First, we explore the distance to which we will be able to detect GC streams with  $C_{\rm err} > 15$ for a 1 hour exposure ($Z087 <$ 28.54) with WFIRST. We first use the normalized luminosity function shown in the bottom panel of Figure \ref{fig:iso_cfht}, to obtain the number of resolved stars WFIRST will observe in GC streams at various distances. After moving the stars in the luminosity function to farther distances, in Figure \ref{fig:distance} (left panel) we show the number of resolved stars in a Pal 5-like stream (blue solid line), a 5 times more massive stream (blue dotted line) and a 10 times more massive stream (blue dashed line) as a function of distance to external galaxies. The vertical lines show the distances to four well-known galaxies in which it could be interesting to search for GC stellar streams. The right panel of Figure \ref{fig:distance} shows the same calculation for a 1000 sec. exposure time with WIFRST, which reaches a limiting magnitude of $Z087 <$ 27.15. See {\it Appendix} \ref{sec:appendixA} for details on the number of resolved stream stars in a 1000 sec. WFIRST exposure. 

We evaluate the minimum number of stars needed in the Pal 5-like, 5 $\times$ Pal 5-like and 10 $\times$ Pal 5-like stream to obtain $C_{\rm err} > 15$ at each distance for a 1 hour exposure (Figure \ref{fig:distance}, left) and a 1000 sec. exposure (Figure \ref{fig:distance}, right). In particular, we calculate the average number of stars in the background covering the same area as the stream at each distance. To assess the stellar density in the background, we first use the simulated M31 fields described in Section \ref{sec:WFIRST}, and we keep track of which stars have magnitudes above the limiting magnitude of WFIRST for a 1 hour and 1000 sec. exposure at each distance. Note that both the number of stars in the background and the number of stars in the streams decline with distance. For simplicity, we evaluate $C_{\rm err}$ using all metallicities, and we assume that the mock-streams are located at $R_{GC} = 35$ kpc to fix the areas of the mock-streams with three different masses. This analysis gives us an estimate of the threshold number of stars needed to detect the streams at each distance assuming perfect separation (i.e. no contaminating background galaxies: see red lines). 
 
The solid red line shows the number of stars needed in a Pal 5-like stream, in order for $C_{\rm err}$ = 15, and the dotted and dashed lines show the same $C_{\rm err}$ = 15 thresholds for the 5 $\times$ Pal 5-like and 10 $\times$ Pal 5-like streams, respectively. If the red lines are above the blue line of the respective mock-stream, C$_{err} <$ 15 and the stream is not easily observed; if the red line is below the respective stream, C$_{err} >$ 15, and the stream is easily observed at that distance.  The shape of the red lines differ from the shape of the blue lines, as the M31 background (red lines) has multiple populations of stars with various ages and metallicities (see Section \ref{sec:WFIRST}), where the blue lines only represent the stellar population of a Pal 5-like cluster.

From Figure \ref{fig:distance}, for the 1 hour WFIRST exposure time (left panel), assuming perfect star/galaxy separation, we conclude that Pal-5 like streams should stand out clearly against the stellar halo background stars if the stream is located in a galaxy $<$ 1.8 Mpc from the Milky Way (see red star where the solid blue and red lines cross). The 5 and 10 times  more massive Pal 5-like streams, will have C$_{err} >$ 15 if they are residing in galaxies $<$ 6.2 and $<$7.8 Mpc from the Milky Way (see red stars where the dashed and dotted blue and red lines cross). 

Similarly, for the 1000 sec. WFIRST exposure time (Figure \ref{fig:distance}, right), we find that the Pal-5 like mock-stream located at $R_{GC}$ = 35 kpc will have C$_{err} =$ 15 at the distance of M31 and should therefore not stand out clearly against the stellar halo of a respective host galaxy much further away than M31 (see also {\it Appendix} \ref{sec:appendixA}). The 5 and 10 $\times$  more massive Pal 5-like streams, will have C$_{err} >$ 15 if they are residing in galaxies $<$ 2.8 and $<$3.0 Mpc from the Milky Way, respectively (see red stars where the dashed and dotted blue and red lines cross) and should be easily detectable as over densities out to those distances.

Using the simulated WFIRST M31 field including contaminating galaxies, we re-calculate the average number density of contaminating objects (stars and now also galaxies: see Section \ref{sec:contamination}) in the background at several distances, covering the same areas as the mock-streams located at $R_{GC} = 35$ kpc. In particular, we re-compute the red lines in Figure \ref{fig:distance} using the new background field, and we keep track of which objects have magnitudes above the limiting WFIRST magnitude for a 1 hour (left) and 1000 sec. (right) exposure at various distances (see gray lines in Figure \ref{fig:distance}). 

As there are now more objects in the background, we need a higher number of mock-stream stars, in order for the thin streams to stand out against the background (i.e for C$_{err} = 15$, see Equation \ref{eq:cerr}). In particular, we found that for a 1 hour exposure using the backgrounds which include contaminating galaxies, the distance limit to observe thin streams in external galaxies will be smaller (see black circles where gray lines intersect blue lines in Figure \ref{fig:distance}, left). For a Pal 5-like stream with a 1 hour WFIRST exposure, the stream should stand out clearly against the stellar halo background (i.e. C$_{err} >$ 15) if the stream is located in a galaxy $<1.1$ Mpc from the MW. The 5 and 10 times more massive Pal 5-like streams, will have C$_{err} >$ 15 if they are residing in galaxies $<1.7$ and $<3.5$ Mpc from the Milky Way, respectively. 

Similarly, for the 1000 sec. exposure including contaminating galaxies (see gray lines in Figure \ref{fig:distance}, right) the distance limit to which we can observe thin streams decreases. In fact, the Pal 5-like stream will not be observable beyond M31 as the gray solid line is above the blue solid line at all distances. An extrapolation of the solid gray line to smaller distances than M31, however, indicates that Pal 5-like streams could be detectable out to $\sim$400 kpc with a 1000 sec. exposure. Using a 1000 sec. exposure with WFIRST, the 5 and 10 times more massive streams will be observable out to galaxies $<$0.8 and $<$2.0 Mpc from the Milky Way, respectively.  Thus, if the goal of a future WFIRST survey is to find GC streams in external galaxies, it might be better to target four times fewer galaxies and observe them for four times as long, as the GC streams will then stand out more clearly against the background stellar halos. Although even within a volume of 400 kpc surrounding the Milky Way, there are $>$47 dwarf galaxies (\citealt{kara19}, see more details in Section \ref{sec:kara}).
 
As we might be able to separate stars and galaxies better with machine learning techniques, the gray lines in Figure \ref{fig:distance} are meant to represent  a ``worst case" scenario. The true distance limit to which WFIRST will be able to observe thin, globular cluster streams is likely somewhere in between the black circles and red stars shown in Figure \ref{fig:distance}. To summarize, for a 1 hour WFIRST exposure, depending on the contaminating background, Pal 5-like streams should be observable out to between 1.1 - 1.8 Mpc, 5 times more massive streams should be observable out to between 1.7 - 6.2 Mpc and 10 times more massive streams should be observable out to between 3.5 - 7.8 Mpc. It is possible that even more massive globular cluster streams exist in these galaxies, and our estimates are therefore conservative.

Note that the contaminating background sources will also depend on the exact location and distance to the external galaxy. Furthermore, our background estimates are based on M31's stellar halo at $R_{GC} = 35$ kpc, and the halos of other galaxies, and in particular stellar densities, can be different (most notably for low mass galaxies). Estimating the backgrounds of galaxies closer to us than M31, would require adding more stars to M31's halo, because there would be more stars brighter than the limiting magnitude. As the galaxies closer to us than M31 are low mass galaxies, and therefore have stellar halos which differ substantially from M31's halo, we do not estimate the background densities of these.

\subsubsection{Crowding}
\label{sec:crowding}
The results presented in Figure \ref{fig:distance} are only valid if all stars will be resolved with WFIRST at each distance for the given number densities of the stellar streams. To test this for all mock-streams, which span various areas, we calculate the number density of stars. In particular, we calculate the number of stars that can be resolved with WFIRST at a given distance and divide this by the area of each stream in arcsec$^2$ at this given distance. As the stream is moved to a greater distance, the number of resolved stars decrease due to WFIRST's limiting magnitude (see Figure \ref{fig:distance}), and the area of the streams decrease. Additionally, the area of the nine different streams vary due to the change in length and width of the streams depending on their location in their host galaxies and their masses (see Table \ref{tab:Pal5}).

Using the number density above, we can test whether there will be crowding in WFIRST's pixels (i.e. more than 1 star per pixel), by comparing the number density to the pixel size of WFIRST, which is $0.11\arcsec \times 0.11\arcsec$. To do this, we multiply the number of stars per arcsec$^2$ in each stream at each distance by the area of a WFIRST pixel in arcsec$^2$.  The mock-stream for which crowding would have the largest effect, is for the most massive 10 $\times$  Pal 5-like stream at small galactocentric radii ($R_{GC}$ = 15 kpc), where its area covers the smallest region. 

We find that even for this most extreme case, crowing is not an issue. In particular, the number of stream stars per WFIRST pixel increases with distances up to $\sim$5 Mpc by $\sim$1.5 dex, and then plateaus at less than 0.01 stream stars per WFIRST pixel. For comparison, for the Pal 5-like stream at the largest galactocentric radius ($R_{GC}$ = 55 kpc), the number of stream stars per WFIRST pixel never exceeds $10^{-4}$. Thus, crowding is not a concern for detecting thin cold, streams with WFIRST.

\subsection{Cold streams in external galaxies: Integrated light}
\label{sec:integrated}
We can also search for thin stellar streams using integrated light. Current telescopes such as HSC (\citealt{miyazaki12}) and future surveys as WFIRST's imaging program (\citealt{spergel13}), LSST (\citealt{ivezi08}) and Euclid (\citealt{racca16}) are ideal for integrated light searches for thin, stellar streams. 

To address whether we should find streams in external galaxies in integrated light, we compute the surface brightness of the Pal 5-like, 5 and 10 times  more massive streams at three different galactocentric radii ($R_{GC}$ = 15  kpc, $R_{GC}$ = 35 kpc and $R_{GC}$ = 55 kpc, see Table \ref{tab:Pal5}). 
As there is scatter in the IMF sampler from the luminosity function, we repeat the sampling from the luminosity function 100 times, to estimate the standard deviation and the mean surface brightness value.  
See all values in Table \ref{tab:Pal5}. 

While the surface brightness of a stream is independent of the distance to its host galaxy, the surface brightnesses will change depending on the telescope bands, however.
We calculate the surface brightnesses of the mock-streams in WFIRST, LSST, Euclid and HSC bands by downloading isochrones from the PARSEC evolutionary tracks (\citealt{bressan12}) assuming Pal-5 like properties ([Fe/H] = $-1.3$, Age = 11.5 Gyr). The mock-streams are brightest if they were located at small galactocentric radii (the stream areas are smaller) and if they are massive (the stream surface densities are higher). The mock-streams are brightest in the F-band for WFIRST, $y$ -band for LSST, $z$-band for HSC and $H$-band for Euclid. 

\begin{figure}
\centerline{\includegraphics[width=\columnwidth]{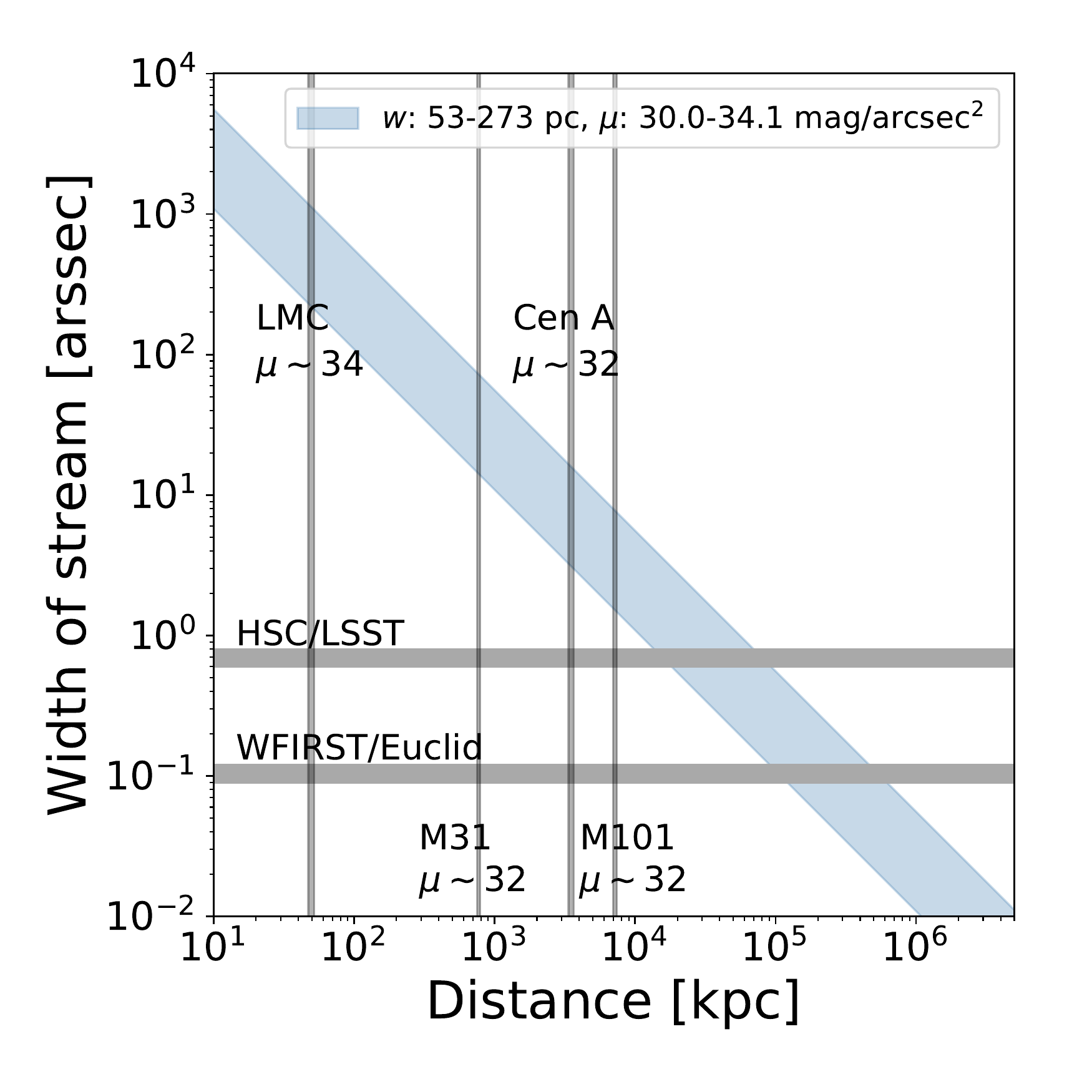}}
\caption{
The blue shaded region shows the angular width in arcsec for the nine mock stellar streams (53 - 273 pc) as a function of distance to external galaxies. The stream widths are calculated at three different galactocentric radii in a host galaxy with a mass-profile similar to M31. The stream is thinnest at smaller $R_{GC}$ and for the lowest mass mock-stream (see Section \ref{sec:length}). The vertical lines show the distances to four well-known galaxies which are labeled with their typical halo surface brightnesses in units of mag arcsec$^{-2}$. The horizontal lines show the angular resolution of four different current (HSC) and future (LSST, WFIRST, Euclid) surveys. The intersects between the blue shaded region and the horizontal lines show the distance limits of where the stream widths can be resolved with various telescopes. From left to right, the first intersect between the blue shaded region and the horizontal lines demonstrates the distance to which a Pal 5-like stream can be resolved, and the second intersect shows the distance to which the width of a 10 times more massive Pal 5-like stream can be resolved. Note, however, the low surface brightnesses of the mock stellar streams (see figure legend and Table \ref{tab:Pal5}).} 
\label{fig:int}
\end{figure}

From the brightest ($F184$) to the faintest ($R062$) WFIRST bands, the surface brightnesses are between $\mu =$ 30.0-34.1 mag arcsec$^{-2}$ for all mock-streams.  The fact that these surface brightnesses are so faint will make the mock-streams extremely hard to detect in integrated light. The surface brightness for the LSST $g$ to $y$-bands are even fainter with $\mu =$ 31.8-36.1 mag arcsec$^{-2}$. Similarly, for HSC,  $\mu =$ 31.9-35.1 mag arcsec$^{-2}$ from the B to z-bands. For Euclid, the streams will be slightly brighter: $\mu =$ 29.3-31.7 mag arcsec$^{-2}$, however the wide field Euclid survey should not probe magnitudes this faint ($\mu_{\rm limit} \sim$ 28.7 mag arcsec$^{-2}$), but the deep field might ($\mu_{\rm limit} \sim$ 29.7 mag arcsec$^{-2}$).  Thus, the streams will be brightest in WFIRST and Euclid bands, but will still be quite faint. 

As the case for resolved streams, more sophisticated techniques than ``by eye" identification may substantially help the detectability of stellar streams in external galaxies in integrated light. Additionally, a younger, more metal rich cluster than Pal 5 would have more bright stars and should therefore have a higher surface brightness. To test what surface brightness the mock-streams would have if they originated from a young, metal rich cluster, we compute the surface brightness for mock-streams generated with an isochrone with [Fe/H] =  0, Age = 500 Myr (see Section \ref{sec:WFIRST}). We emphasize that it would be very difficult to form a stream with a significant length only 500 Myr after formation of the cluster. Nevertheless, we use these isochrones to demonstrate an extreme case. 

For the isochrones with Age = 500 Myr and  [Fe/H] =  0, we found a range of $\mu =$ 26.8-30.7 mag arcsec$^{-2}$ in the WFIRST $F184$ to $R062$-bands for the three different mass mock-streams at all galactocentric radii. While it might be feasible to detect such a stream in integrated light against the background of a given galaxy, we conclude that such a stream will be difficult to detect, as the cluster needs to be very young and yet have formed a stream.

Another component of observing thin stellar streams in integrated light is their widths. The angular width of streams decreases with distance to the external host galaxy. To determine whether we would resolve GC stellar streams with current and upcoming telescopes, given the telescopes resolution limits, in Figure \ref{fig:int} we plot the range of widths for our nine mock-streams (blue shaded band) as a function of distance to the host galaxy. Recall that smaller galactocentric radii and lower mass streams, yield narrower streams.

The vertical lines in Figure \ref{fig:int} show the distances to four nearby galaxies, which could be targeted for thin stellar stream observations. We have labeled the vertical lines by the typical surface brightness of the stellar halos for these systems. In particular, the Large Magellanic Cloud (LMC) has a surface brightness of $\mu \sim 34$ mag arcsec$^{-2}$ beyond 20 kpc (\citealt{nidever18}), M31 has a typical surface brightness of  $\mu \sim$ 32 mag arcsec$^{-2}$ beyond 40 kpc (\citealt{ibata07}, fig. 42), Cen A has  $\mu \sim $ 32 mag arcsec$^{-2}$ (\citealt{crno16}), and M101 has  $\mu \sim 32 $ mag arcsec$^{-2}$ beyond 40 kpc (\citealt{dokkum14}, fig. 2). See the range of the Pal 5-like mock-stream widths and their surface brightnesses in the legend of Figure \ref{fig:int} and in Table \ref{tab:Pal5}. The horizontal lines show the resolution limitations of HSC, LSST (limited by seeing: 0.7$\arcsec$) and WFIRST, Euclid (limited by the pixel scale: 0.11$\arcsec$ for WFIRST: \citealt{spergel13} and a pixel scale of 0.1$\arcsec$ for Euclid: \citealt{racca16}). 

From Figure \ref{fig:int}, we conclude that Pal 5-like streams can be resolved by HSC and LSST to distances of ${\sim} 20$ Mpc, and that 10 times  more massive streams should be resolved out to $ \sim$100 Mpc (see intersect of HCS/LSST horizontal line with upper and lower limit of blue shaded band). Similarly, WFIRST and Euclid will be able to resolve a Pal 5-like stream out to $\sim$100 Mpc, and a ten times more massive stream out to $\sim$600 Mpc (see intersect of WFIRST/Euclid horizontal line with upper and lower limit of blue shaded band). At a distance of $\sim$ 600 Mpc, however, the length of a 10 times  more massive stream than Pal 5 would only be ${\sim} 6\arcsec$ (and the width is ${\sim} 0.11\arcsec$). 

As noted above, we caution that unless the thin globular cluster streams in the external galaxies originate from very young, metal rich clusters, the very low surface brightness of Pal 5-like streams ($\mu =$ 30.0-34.1 mag arcsec$^{-2}$ for WFIRST) make them very difficult to detect in integrated light in any nearby galaxy.

\section{Discussion}
\label{sec:discussion}

\subsection{External galaxies near the Milky Way}
\label{sec:kara}
Based on our findings in Section \ref{sec:results}, a 1 hour WFIRST exposure will be able to detect Pal 5-like streams in resolved stars in galaxies $\sim$1.1 - 1.8 Mpc away from the Milky Way, depending on the exact galaxy contamination fraction. We encourage observers to look for these thin streams in the future. 

\citet{kara19} recently presented the distribution of galaxies within 11 Mpc of the Milky Way, 5 out of 6 of which are dwarf galaxies (see their figure 4). Using the data presented in \citet{kara19}, we find that there are 109 - 119 galaxies located between 1.1 and 1.8 Mpc from the Milky Way, where 1.8 Mpc refers to the perfect star/galaxy separation (see red solid line in Figure \ref{fig:distance}), and where 1.1 Mpc is the distance limit estimated when including contaminating galaxies as described in Section \ref{sec:contamination} (see also gray solid line in Figure \ref{fig:distance}). If a Pal 5-like stream is located in either of the 109 - 119 galaxies, WFIRST should be able to detect it in resolved stars. All of these galaxies, except for M31, are dwarfs, and the most luminous dwarfs within this region are M32, M33 and the LMC. The number of observable streams will depend on the GC populations of dwarfs (see e.g. GCs in Fornax: \citealt{shapley38}, \citealt{hodge61a} and the LMC: \citealt{hodge61b}). The exact properties of the streams will further depend on the strength of the dwarfs' tidal fields. 

Similarly, we estimated that the 5 times more massive Pal 5-like streams are easily detectable in external galaxies within $\sim$1.7 - 6.2 Mpc of the Milky Way (see Section \ref{sec:resother}). 
Within these distance limits, there are 118 - 493 galaxies (\citealt{kara19}, I. Karachentsev, {\it private communication}). Only 4\% of these have luminosities, $L > 5 \times 10^9$ $L_{\odot}$. Similarly, for the 10 times more massive streams, we estimated that the distance limit to external galaxies was  $\sim$3.5 - 7.8 Mpc (see dashed gray lines in Figure \ref{fig:distance}). There are between 199 - 667 galaxies within these limits, and, again, only 4\% of these are more luminous than  $L > 5 \times 10^9$ $L_{\odot}$.

We emphasize that the distance limits to external galaxies presented in Figure \ref{fig:distance} could be larger with stricter color cuts (i.e. metallicity cuts), and as the GCs could be younger and therefore brighter than Pal 5.  Moreover, our constraint that the mock-streams stand out with respect to background stellar halos if C$_{err} >$ 15 could be too conservative. A smaller value than C$_{err} =$ 15 might enable us to include more massive galaxy hosts to surveys searching for thin GC streams (see also the discussion in Section \ref{sec:color}).

\subsection{Finding thin globular cluster streams}
\label{sec:color}
There are additional techniques we can apply which can facilitate detections of thin streams to greater distances. Globular clusters have metallicity spreads from [Fe/H]$ = -2.5$ - 0 (\citealt{harris96}). Depending on the properties of the stellar halo of interest, the cluster can stand out more against the background stellar halo of their host galaxy than illustrated in Figure \ref{fig:M31_wfirst} if the clusters have lower metallicities than Pal 5. PAndAS and WFIRST provide photometric metallicities and in Figure \ref{fig:M31_pandas} and \ref{fig:M31_wfirst} we have therefore implicitly used the color info to apply a ``metallicity cut" (color-cut). 
However, several colors are available for WFIRST and more cuts in color-color diagrams could help detect fainter streams (see e.g. \citealt{shipp18}). 

In addition to color cuts, we might be able to detect more streams in external galaxies by doing orbit searches using algorithms such as ``stream-finder" (e.g. \citealt{malhan18}, \citealt{ibata19}). This might enable us to find thin stellar streams at larger distances than suggested in Figure \ref{fig:distance}. However, the lack of kinematic information could complicate these searches. Another method to improve detectability is to use a matched filter technique by applying a smoothing criteria matching the estimated width of the stream at a given distance to the external galaxy of interest. Even without kinematics, we can use the fact that streams are continous and search for them using machine vision algorithms such as the ``Rolling Hough Transform" (see \citealt{clark14}). 

In Section \ref{sec:lum}, we noted that using a 1 hour exposure, WFIRST will not be able to resolve stars in a Pal 5-like stream to greater distances than 38 Mpc, as the limiting magnitude of WFIRST ($Z087 = $ 28.54) would be brighter than the brightest stars in Pal 5. %
The observations of the ``Maybe Stream" presented in \citet{abraham18}  were carried out with one HST orbit, and they reached a limiting magnitudes of AB ${\sim} 26.5$ (Abraham, private communication). The tip of the red giant branch of Pal 5 would be below the detection limit at this distance, as the isochrone shifts by ${\sim}$14.6 magnitudes if the stream is at 20 Mpc instead of 23.5 kpc (i.e. $d_{mod}$ =  31.51 instead of $d_{mod}$ = 16.86, see isochrone in the bottom left part of Figure \ref{fig:iso_cfht}). Therefore, our results indicate that it is unlikely that the reported ``Maybe Stream" at 20 Mpc (\citealt{abraham18}) hosts resolved stars from the remnant of a Pal 5-like globular cluster. It is possible that \citet{abraham18} are observing the stream in integrated light (see Figure \ref{fig:int}), however the width of the ``Maybe Stream" stream is only 0.1$\arcsec$ at 20 Mpc, which indicates the stream is only 10 pc wide. Thus assuming 10 pc is representative of the full width of the ``Maybe Stream", the stream would be of lower initial mass than Pal 5's cluster. As a comparison, the thin MW streams, GD1 and Pal 5 range in widths from 30-60 pc (\citealt{price18}) and 40-80 pc (Bonaca et al., {\it in prep.}), respectively. It therefore seems unlikely that a low mass cluster would produce such a long, bright stream. One possibility is that the stream's progenitor cluster was much younger than Pal 5 and therefore leaves the ``Maybe Stream" with more bright stars (see  Section \ref{sec:integrated}). It is also possible that the  ``Maybe Stream" could be closer to us than its nearby galaxies, but that would not explain the issue of its thinness. Deeper observations might help elucidate these remaining puzzles. 

\subsection{Gaps in cold stellar streams in external galaxies}
\label{sec:gaps}
The gravitational interactions between dark matter subhalos and stellar streams provide an intriguing method for probing the dark matter subhalo power spectrum and thereby setting limits on the nature of the dark matter particle (see e.g. \citealt{erkal16}, \citealt{bovy17}, \citealt{price18}, \citealt{bonaca19}). \citet{garrison17} showed that the presence of dark matter subhalos should be suppressed substantially within the orbit of Pal 5 if baryonic disks are included in dark matter only simulations of galaxies.  Despite this fact, gaps and irregularities have been reported in Pal 5 (see e.g. \citealt{erkal17}, Bonaca et al., {\it in prep.}). These irregularities likely arise to due to Pal 5's prograde orientation with respect to the disk of the Galaxy, as Pal 5 will more likely be affected by molecular clouds in the disk (\citealt{amorisco16}), torques from the Galactic bar (\citealt{hattori16}, \citealt{erkal17}, \citealt{pearson17}) or interactions with spiral arms (\citealt{banik19}).

The MW GC stellar stream, GD1 (\citealt{grillmair06}), has an orbit which probes a similar region of the Galaxy as Pal 5 (GD1's $r_{peri} \sim 14$ kpc, GD1's $r_{apo} \sim 26-29$ kpc, \citealt{koposov10}). The orientation of its orbit, on the other hand, is retrograde with respect to the disk of the Galaxy and its pericentric distance is larger than Pal 5's. This makes GD1 a cleaner laboratory for searching for potential past interactions with dark matter subhalos. Interestingly, the recent detection of an under density and a ``spur" in GD1 using data from Gaia DR2 (\citealt{gaiadr2}) can be interpreted as an interaction with a dense substructure (\citealt{price18}, \citealt{bonaca19})

Our work shows that WFIRST should detect many thin stellar streams in resolved stars in nearby galaxies. This will open up the possibility of exciting indirect detections of dark matter through density distortions and gaps in thin stellar streams. In external galaxies, we have the advantage of being able to select galaxies without spiral arms and bars, limiting the possible stream perturbers. As discussed in Section \ref{sec:kara}, most of the galaxies in the Local Volume are low mass galaxies (\citealt{kara19}), which could be ideal stream hosts. 

Furthermore, we will  have the opportunity to look for streams at greater galactocentric distances, where more subahlos should reside (\citealt{garrison17}), and we can potentially build up statistical samples to measure subhalo properties as a function of host mass. 
We will still have to exclude interactions with globular clusters and satellites (see e.g. \citealt{bonaca19}), which will be more difficult due to the lack of kinematic information. However, we hypothesize that vastly increasing the number of thin stellar streams with resolved stars in external galaxies, will be crucial in our quest for an understanding of dark matter. 

\section{Conclusion}\label{sec:conclusion}
In this paper, we have created mock globular cluster stellar streams and investigated the observability of thin, globular cluster streams in external galaxies. We have used the physical properties of the MW stellar stream, Pal 5, as our reference stream, and we have explored GC streams with a maximum of ten times the mass of Pal 5's stream. Our estimates are conservative, as more massive GC streams can exist. Based on our findings we make the following conclusions:

\begin{itemize}
\item[1.]  For a survey with a 1 hour (1000 sec.) exposure, including contaminating galaxies, WFIRST will be able to detect old, metal poor globular cluster stellar streams in resolved stars to 3.5 Mpc (2.0 Mpc). This volume contains 199 (122) galaxies (\citealt{kara19}), of which 7 (1) are more luminous than 0.1$\times {\rm L}_{\rm MW}$. The exact number of streams will depend on the globular cluster population in the surveyed galaxies.
\item[2.] Assuming perfect star/galaxy separation, for a 1 hour (1000 sec.) exposure, WFIRST will be able to detect old, metal poor globular cluster stellar streams in resolved stars to distances of 7.8 Mpc (3 Mpc), a volume which contains 667 (150) galaxies, of which 25 (2) are more luminous than 0.1$\times {\rm L}_{\rm MW}$. 
The distance limit for detecting streams in external galaxies is likely somewhere in between the limits presented in 1. (perfect star/galaxy separation) and 2. (conservative star/galaxy separation).
\item[3.]  Current and future imaging surveys, using integrated light, can resolve thin stellar streams out to distances of $\sim$600 Mpc, depending on the streams' width and location in their host galaxies. However, the very low surface brightness of the streams (typically $>$30 mag/arcsec$^{2}$) will likely prohibit the detection of such systems, unless the streams originate from much more massive, young, and metal rich clusters than Pal 5. 
\end{itemize}

Our work provides a positive outlook on the future prospects of finding thin stellar streams in resolved stars with WFIRST in external galaxies. We emphasize that detecting these systems in integrated light will be difficult. The thin streams in external galaxies can be used to both indirectly probe the nature of dark matter through gaps in streams, and to map orbit structures in external galaxies and potentially the triaxiality of external dark matter halos (see Yavetz et al., {\it in prep.}). 

\acknowledgements
We thank Rubab Khan, Andrew Dolphin and Eric Bell for their contributions to the WFIRST image simulations and reduction software and catalogs. We also thank Robyn Sanderson, Adrian Price-Whelan, Anil C. Seth, David W. Hogg and David Spergel for insightful discussions, and and we thank Igor Karachentsev for sharing his data on external galaxies. SP's, TKS's and KVJ's work was performed in part during the Gaia19 workshop and the 2019 Santa Barbara Gaia Sprint (also supported by the Heising-Simons Foundation), both hosted by the Kavli Institute for Theoretical Physics at the University of California, Santa Barbara and supported but the the National Science Foundation under grants NSF PHY-1748958. The Flatiron Institute is supported by the Simons Foundation. KVJ’s contributions were inspired by the WFIRST Infrared Nearby Galaxies Survey  collaboration and supported in part by NASA grant NNG16PJ28C through subcontract from the University of Washington and the National Science Foundation under grants NSF AST-1614743.

\software{
   \package{Astropy} ~\citep{astropy:2013, astropy:2018}, 
    ~\package{Matplotlib} ~\citep{Hunter:2007}, 
    ~\package{Numpy} ~\citep{walt2011}, 
    ~\package{Scipy} ~\citep{scipy} 
}

\bibliographystyle{aasjournal}

\appendix
\section{WFIRST exposure times}
\label{sec:appendixA}
Throughout the paper we have used a 1 hour exposure as our guideline for the limiting magnitude of WFIRST. In this {\it Appendix}, we show how some of our results will differ if we instead use the guest observer capabilities of 1000 sec. exposures for WFIRST instead. A 1000 sec. exposure yields a $Z087$-band limiting magnitude of mag $< 27.15$ as opposed to mag $< $ 28.54 for a 1 hour exposure (see \citealt{spergel13}). 

\begin{figure*}
\centerline{\includegraphics[width=0.7\textwidth]{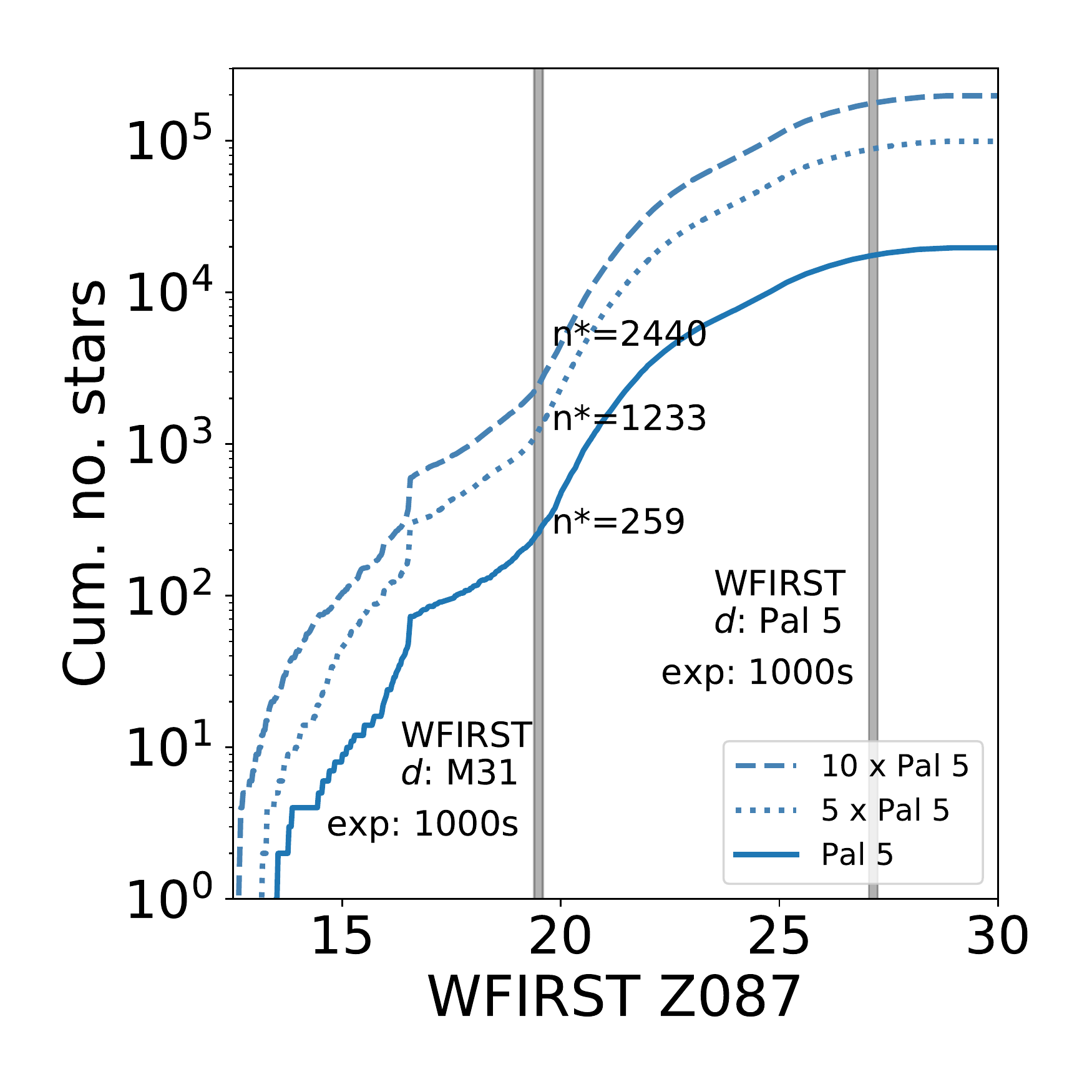}}
\caption{The cumulative number of stars in a Pal 5-like stream (solid line), a 5 $\times$  more massive Pal 5-like stream (dotted line), and a 10 $\times$  more massive Pal 5-like stream (dashed line) for a given limiting $Z$-mag. The vertical lines show the limiting magnitude of WFIRST for a 1 hour exposure ($Z087 < 27.15$) at the distance of M31 (i.e. shifted by 7.66 magnitudes from Pal 5's current location), and at the distance of Pal 5 in the Milky Way. We indicate the amount of stars WFIRST should be able to observe for the Pal 5-like, 5  $\times$ Pal 5-like and 10 $\times$ Pal 5-like stream in M31 (see n$^*$). 
} 
\label{fig:A1}
\end{figure*}

In Figure \ref{fig:A1}, we recompute the cumulative number of stars WFIRST would observe in a Pal 5-like, 5 $\times$ Pal 5-like and 10 $\times$ Pal 5-like stream (see Figure \ref{fig:iso_cfht} for details) for a 1000 sec. exposure with WFIRST as a function of  $Z087$-band limiting magnitude. From the Figure, we conclude that we lose $\sim$5 times the amount of stars for a 1000 sec. exposure as compared to the amount of observable resolved stars for a 1 hour exposure (see Figure \ref{fig:iso_cfht}). 

\begin{figure}
\centerline{\includegraphics[width=1\textwidth]{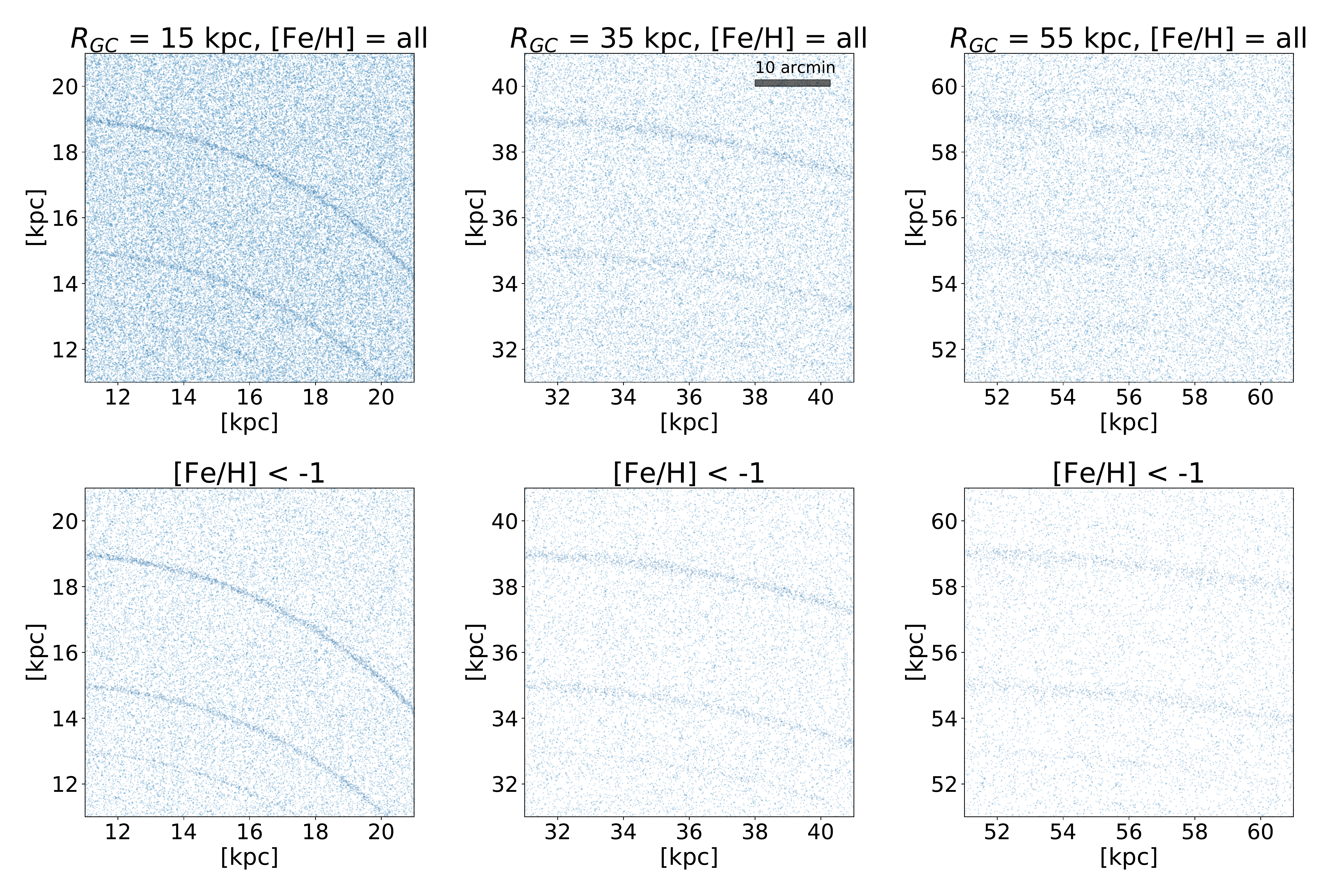}}
\caption{This figure shows the same panels as Figure \ref{fig:M31_wfirst}, but now we determine the number of stars in each mock-stream by summing up the cumulative number of stars in the streams at the limiting magnitude of WFIRST ($Z087 < 27.15$) at the distance of M31. We use a point size which is 10 times smaller than in Figure \ref{fig:M31_pandas} to avoid crowding of the background. The black bar in the top middle panel shows the scale of 10 arcmin for reference. There are 2440 $\pm$ 165 stars in the 10 $\times$ more massive stream, 1233 $\pm$ 86 stars in the 5 $\times$ more massive stream and 259 $\pm$ 17 stars in Pal 5-like stream (see Figure \ref{fig:A1}).  All stream stars are assumed to have [Fe/H] $= -1.3$, thus all stream stars are resolved in each panel. At each $R_{GC}$, we have updated the width and lengths of the streams based on the tidal field they experience at these distances (see Section \ref{sec:length}). Additionally, we have updated the number of stars in each background field to illustrate what WFIRST will observe in M31 for a 1000 sec. exposure.  
The 5 and 10 $\times$ Pal 5-like mock-streams are clearly visible in each panel, and hints of the Pal 5-like stream are visible in the bottom row after applying metallicity cuts to the backgrounds.} 
\label{fig:A2}
\end{figure}

As we go to brighter limiting magnitudes, the amount of stars in the background of M31 will also decrease vastly. We recompute Figure \ref{fig:M31_wfirst} using the updated amount of stars in each stream (see Figure \ref{fig:A1}). We estimate the stellar halo backgrounds as described in Section \ref{sec:WFIRST}, but now with a limiting magnitude $< 27.15$ instead of $< $ 28.54. In Figure \ref{fig:A2}, we show the results of injecting three mock-streams in each 10 $\times$ 10 kpc region, using a limiting magnitude of $< 27.15$ to estimate the number of stars in the background and in each mock-stream.  
We conclude that using the guest observer capabilities of WFIRST in M31, yields a similar (but slightly better) result for resolving GC streams as currently possible with the PAndAS data (see Figure \ref{fig:M31_pandas}). In particular, we find that $C_{err} \sim 15$ (see Section \ref{sec:Cerr}) for the Pal 5-like stream in the top row, and that the Pal 5-like streams are more apparent in the bottom row, where $C_{err} > 20$. For the 5 and 10 more massive streams, $C_{err}$ is much larger than 15, and the streams will be clearly visible against the background.

\end{document}